\documentclass[prb,twocolumn,showpacs]{revtex4}
\usepackage{graphicx,epsf}
\newcommand{\fett}[1]{\mbox{\boldmath $#1$\unboldmath}}
\begin{document}
\title{Spin-glass phase of cuprates}
\author{N. Hasselmann$^{1,2}$, A.~H.~Castro Neto$^{3}$, and 
C.~Morais Smith$^{4}$} 
\affiliation{$^1$ Max-Planck-Institut f\"ur Physik komplexer Systeme, N\"othnitzer Str.~38, 01187 Dresden, Germany \\
$^2$ Institut f\"ur Theoretische Physik, Universit\"at Frankfurt, Robert-Mayer-Str.~8, 60054 Frankfurt, Germany \\
$^3$ Department of Physics, Boston University, 590 Commonwealth Ave, Boston, MA, 02215 \\
$^4$ D\'epartement de Physique, Universit\'e de Fribourg, P\'erolles, 
CH-1700 Fribourg, Switzerland.}

\date{\today}

\begin{abstract}
We investigate a phenomenological model for the spin glass phase of 
$\rm La_{2-x}Sr_xCuO_4$, in which it is assumed that holes doped
into the $\rm CuO_2$ planes localize near their Sr dopant, where
they cause a dipolar frustration of the antiferromagnetic environment.
In absence of long-range antiferromagnetic order, the spin system can
reduce frustration, and also its free energy, by forming a state with
an ordered orientation of the dipole moments, which leads to the appearance
of spiral spin correlations. To investigate this model, a non-linear
sigma model is used in which disorder is introduced via a randomly
fluctuating gauge field. A renormalization group study 
shows that the collinear fixed point of the model is destroyed
through the disorder and that the 
disorder coupling leads to an additive
renormalization of the order parameter stiffness. Further, the stability
of the spiral state against the formation of topological defects is
investigated with the use of the replica trick. A critical disorder
strength is found beyond which topological defects proliferate. Comparing
our results with experimental data, it is found that for a hole density
$x > 0.02$, i.e. in the entire spin glass regime, the disorder strength
exceeds the critical threshold. In addition, some experiments are proposed
in order to distinguish if the incommensurabilities observed in neutron
scattering experiments correspond to a diagonal stripe or a spiral phase.
\end{abstract}
\maketitle

\section{Introduction}

\subsection{Generalities}
This paper discusses the influence of disorder
on the properties of weakly hole doped 
cuprate materials.
In cuprates, the superconducting state 
emerges through chemical doping of a parent
compound which is insulating and shows  antiferromagnetic (AF) order with
a high critical N\'eel temperature of typically a few hundred Kelvin.
As  a consequence of chemical doping, the 
compounds are intrinsically disordered. Especially at weak doping
concentrations, disorder is known to strongly influence
the behavior of these materials. 
This is evident in the simplest cuprate superconductor, 
$\rm La_{2-x}Sr_xCuO_4$, 
where the superconducting phase emerges via doping directly from a 
low temperature spin glass (SG) phase. Recently, glassy characteristics
were detected even inside the superconducting phase (see Ref.\ \cite{julien}
for a summary of the available experimental data).

Understanding the very weak doping regime of cuprates,
the insulating AF and SG regime, should be
relatively simple. 
This optimism is based on the belief that this regime
is dominated by the behavior of isolated holes
in presence of well developed AF moments.
The single hole properties seem now to be quite well understood
and early theories of high temperature superconductivity
were constructed from these one-hole wave functions. Shraiman and Siggia 
\cite{shraiman88b,shraiman89a}
proposed a theory of interacting hole-quasiparticles based on the
one-hole picture
and predicted the formation
of spiral correlations with a pitch proportional to the hole density.
Experiments have to date however not found any evidence of such spiral
correlations inside the superconducting phase.
The pairing mechanism
suggested by this semiclassical picture, a dipole-dipole
interaction between holes mediated by soft spin 
waves,\cite{shraiman89b,kuchiev93} has, perhaps unfairly, 
received scant attention of late. 
A potential weakness of the approach is the semiclassical 
treatment of spins (large $S$), which implies the assumption of 
a large AF correlation length, whereas in the superconducting
phase the spins are believed to form some kind of quantum disordered
spin liquid. The scattering of holes by spin excitations would then
be qualitatively different at large scales. 
However, while the semiclassical theory is formulated for large scales,
the structure and energy of the resulting two-hole bound state is
determined by the shortest cutoff in the system\cite{kuchiev93}, where
AF correlations are still intact. 
Furthermore, the correlation length can be substantial even in 
superconducting samples, e.g. it exceeds 200 \AA\ in the stripe compound
La$_{1.45}$Nd$_{0.4}$Sr$_{0.15}$CuO$_4$.\cite{wakimoto03} 
Thus, the pairing mechanism suggested by the semiclassical picture
may hide some truth despite the absence of long range order.

While a semiclassical approach to the superconducting
regime may or may not be valid, at 
sufficiently low hole concentrations, where static AF correlations
are still dominant, i.~e. in the SG and AF phase,
a semiclassical treatment of spins is certainly justified. However,
at these low densities,
where the system is still a Mott insulator, screening is very poor
and long-range Coulomb interaction leads to a strong
disorder potential which must be taken into 
account.
Here we discuss a model in which
the entire charge distribution is assumed to be quenched.
Each hole, localized close to an ionized dopant, 
is assumed to
produce a long-ranged dipolar-shaped frustration of the AF, 
similar to the one known to be produced by delocalized holes.
A polarization of the dipole moments then implies
the appearance of spiral correlations. 

It is known
that the spiral state described by Shraiman and Siggia, if one
ignores disorder,
is unstable toward a local enhancement
of the spiral pitch. This instability arises from the fermionic
susceptibility of the holes and may signal an instability towards
charge density formation or phase separation.\cite{auerbach}
However, if the holes are quenched this instability is suppressed.
Therefore, disorder takes a prominent role in the creation of a 
spiral state.

We here develop a renormalization approach 
for disordered spiral phases, where
we study the scaling of the spin stiffness and of the disorder.
The importance of topological defects of the spiral texture
is analyzed and their relevance for the physics of
the spin glass phase is discussed.

\subsection{Undoped and weakly doped cuprates}

Undoped $\rm La_2CuO_4$ is a charge transfer insulator with an
antiferromagnetically ordered ground-state. 
It is well described by a simple square
lattice spin-$1/2$ Heisenberg model,
\begin{equation}
\label{sqlh}
H_H =  J \sum_{\left<ij\right>} {\bf S}_i \cdot {\bf S}_j,
\end{equation}
with the antiferromagnetic exchange $J\sim 1200$K. The sum is over nearest
neighbor pairs of sites and ${\bf S}_i$ are spin-$1/2$ operators.

In the study of magnetism of $\rm La_2 CuO_4$, an
approach based on the quantum-non-linear-$\sigma$ model (QNL$\sigma$M)
has been highly successful. 
It correctly describes the long wavelength hydrodynamic modes (spin waves)
of the Heisenberg model.\cite{chakravarty89} In this continuum model, it is
assumed that the antiferromagnetic correlation length is much larger
than the lattice spacing and the model describes slow fluctuations
of the locally well defined staggered magnetization $\bf n$ 
(with ${\bf n}^2=1$). The QNL$\sigma$M action is
\begin{equation}
\frac{S_{\rm eff}}{\hbar}=\frac{\rho_S}{2 \hbar} \int_0^{\hbar \beta} d\tau
\int d^2 {\bf x} \left\{ \left( \partial_\mu {\bf n}\right)^2 + \frac{1}{c^2}
\left( \partial_\tau {\bf n}\right)^2 \right\}.
\end{equation}
The spin stiffness $\rho_S$ and the spin wave velocity $c$ should be
viewed as phenomenological parameters to be determined either from experiment
or from other techniques such as spin wave theory or numerical simulations. 
The
coupling constant of the model is $g=\hbar c \Lambda /\rho_s $ ($\Lambda$ is
a high frequency cutoff). 
There is a zero temperature quantum phase transition at
$g=g_c\sim 4\pi$ from a phase
with long-range order ($g<g_c$, ``renormalized classical regime'')
to a phase which exists for $g>g_c$ and which is
quantum disordered with only finite spin correlations and no static magnetic
order. 
It is now firmly believed that the $S=1/2$ Heisenberg model described by
(\ref{sqlh}) has $g<g_c$.
Measurements of the correlation length of $\rm La_2CuO_4$
have been fitted extremely well with the QNL$\sigma$M predictions for the
renormalized classical regime.\cite{birgeneau99}

\begin{figure}[hbt]
\begin{center}
\includegraphics[width=5cm]{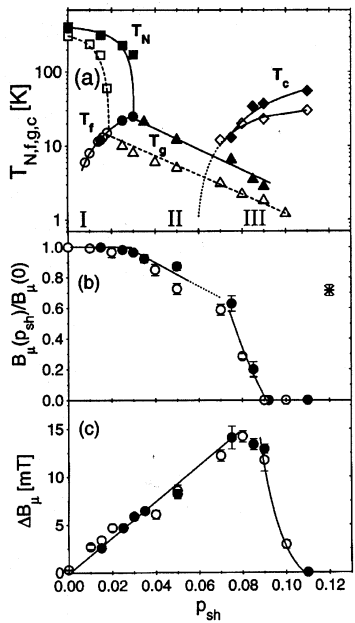}
\end{center}
\caption[]{\label{niederm} Phase diagram as seen by $\mu$-SR, with data 
obtained
from $\rm La_{2-x}Sr_xCuO_4$
(open symbols) and $\rm Y_{1-x}Ca_xBa_2CuO_3$ (closed symbols),
$p_{sh}$ is the hole
concentration. (a) Doping dependence of the N\'eel
temperature $T_N$, freezing transition temperature $T_f$, spin glass
transition temperature $T_g$ and superconducting transition temperature
$T_c$. (b) Normalized average internal field at $T$=1 K.
(c) RMS deviation $\Delta B$ at $T$=1 K. Fig. from Niedermayer
{\em et al.}\cite{niedermayer98}}
\end{figure}

Once holes are added to the CuO$_2$ planes, the magnetism becomes
rather complex. Fig.~\ref{niederm} summarizes the magnetic
phase diagram at weak doping concentrations of 
$\rm La_{2-x}Sr_xCuO_4$ and $\rm Y_{1-x}Ca_xBa_2CuO_4$.\cite{niedermayer98} 
Here, we concentrate on $\rm La_{2-x}Sr_xCuO_4$.
For very small Sr concentration, 
the most dramatic effect is a rapid reduction of $T_N$ with
the complete destruction of long-range order occurring at a 
 critical doping value of roughly $x_g\sim0.02$.
Further, a spin freezing is observed inside the AF phase below
a temperature $T_f$ which scales linearly with the Sr concentration,
$T_f\sim (815 $K$) x$ for $0<x<x_g$. This spin freezing is inferred from
a broad distribution of
extremely slow relaxation times measured with local probes such
as \raisebox{1ex}{\tiny 139}La nuclear quadrupole resonance 
\cite{chou93} (NQR) and muon spin resonance \cite{borsa95} ($\mu$SR).
Surprisingly, while at higher temperatures doping leads to a reduction
of the local staggered moments, at temperatures lower than
about 30 K the staggered moments recover and at zero temperature
they are practically doping independent and 
approach the value of the undoped compound,\cite{chou93,borsa95}
see the middle panel of Fig.~\ref{niederm}.
However, the distribution of staggered moments is broad
at finite doping, with a variance which is again simply linear in $x$,
see  Fig.~\ref{niederm} bottom.\cite{niedermayer98}
Both the recovery of the staggered moments and the broad distribution
of relaxation times are reminiscent of a transverse spin glass state,
in which the transverse spin wave modes of the AF freeze in a 
static but random pattern. 
These are clear signatures of
disorder in the weakly doped AF phase. This is further corroborated
by transport measurements, which show a behavior typical for
random systems.\cite{keimer92} 
At temperatures below $\sim$50 K the conductivity roughly follows
variable range
hopping characteristics while at higher temperatures a thermally activated
conductivity is observed, with activation energies
of about 19 meV.\cite{chen95} This indicates that the
holes localize near the randomly distributed Sr donors.

Both the presence of finite staggered magnetic moments and the
broad distribution of slow relaxation times persist also
above $x>x_g$ where long-range order is destroyed.\cite{niedermayer98} 
Again, there is 
a recovery of the staggered moments at very low temperatures,
although the zero temperature moment is now slightly smaller 
than in the undoped compound. The 
$x$ dependence of $T_f$ follows now roughly a $1/x$ scaling.
The regime $0.02<x<0.05$ is well described as a conventional
spin glass (SG) and shows characteristic non-ergodic
behavior.\cite{wakimoto} The freezing transition temperature
$T_f$  in this regime can thus be identified 
as a SG transition temperature $T_g$.
The fact that staggered moments persist also above $x=0.02$ is 
important and excludes the possibility that
the transition at $x=0.02$ is a disordering
transition driven by quantum fluctuations as described in the QNL$\sigma$M
formulation above. 
It is often argued that 
upon hole doping, the reshuffling of the spins by mobile holes leads
to enhanced quantum
fluctuations of the spins which would eventually drive the
spin system past the quantum critical point of the QNL$\sigma$ model,
driving the AF into a spin liquid phase. 
As the transition
at $x=0.02$ is not followed by a spin liquid phase but rather a 
SG this scenario does not apply for the AF-SG transition.

Only recently, it was found
that the short ranged magnetic order in the SG regime is
incommensurate, with a maximum of the imaginary part of
the susceptibility located at the in-plane wave vector 
($1/2 \pm \delta / \sqrt{2} $, $1/2 \mp \delta / \sqrt{2} $), 
in units of $2 \pi /a $ where $a$
is the Cu lattice spacing.\cite{wakimoto,matsuda,fujita}
Here, $\delta$ is the incommensurability
which roughly follows $\delta\simeq x$.
This
incommensurability has often been interpreted as
diagonal stripe formation, even though no signatures of a charge
modulation were observed. Rather, all experiments point toward
a quenched charge distribution and we thus
argue that a more likely explanation is the formation
of short ranged spiral order.

In $\rm La_{2-x}Sr_xCuO_4$ static AF moments are strong 
for small $x$ and the holes
seem to localize at low temperatures where
transport experiments indicate a relatively weakly bound hole
with a localization length of a few lattice constants. 
Thus, one
might hope to gain considerable insight into these
phases by solving the one-hole problem first and to
proceed from there on. As mentioned in the beginning, the understanding
of the spin-polaron state arising from one hole in an antiferromagnetic
background is by now quite 
mature.\cite{shraiman88b,reiter94,horsch98} 
For the $t-J$ model,
the bottom of the dressed
hole band lies at the zone face centers ${\bf k}_0=(\pm \pi/2,\pm\pi/2)$ and
the bandwidth scales with $J$. Because of the presence of two sub-lattices,
there exists a pseudo-spin degeneracy for each ${\bf k}$ vector. 
An important characteristic
of the hole wave function is that it describes
a long-ranged dipolar distortion of the AF order which arises from
a coupling of the spin current carried by the hole to the magnetization
current of the AF background.\cite{shraiman88b} Relative to the
position of the moving hole, the Fourier transform of the
transverse spin deviations is
then proportional to 
$(\bar{\bf q}_x + \bar{\bf q}_y)/\bar{\bf q}^2$,\cite{reiter94}
where $\bar {\bf q}={\bf q}-(\pi, \pi)$, i.e.~the staggered magnetization
shows a dipolar pattern in real space identical to the one produced by
an isolated ferromagnetic bond, see Sec.~\ref{fbonds}.

The Sr impurity position, located above the center of a Cu plaquette,
has a high symmetry and couples to both sub-lattices in the same
way, so that the pseudo-spin degeneracy
mentioned above should survive also in the bound hole state. 
The bound hole state is a superposition of plane wave states describing
the mobile hole. 
For sufficiently weakly bound holes, we expect
the main weight of the bound hole wave function to remain at wave vectors
close to ${\bf k}_0$ or equivalent positions, and, depending on
the relative phases and the weight of these pockets, dipolar
or quadrupolar frustration is associated with the localized
hole. 
We note that 
dipolar frustration was also suggested by Aharony {\it et al.}
for O doped systems, caused by 
a localization of holes in the O site
with the liberation of  one of the spins from the O $p^6$ 
state,\cite{aharony88} leading to an effective ferromagnetic
coupling for the two Cu spins joint by the O. 
While the microscopic origin of frustration in the Aharony model
is very different from the quantum mechanical one that we assume here,
the phenomenological spin-only model we employ below is not
sensitive to the microscopic details. In either case, the dipole
moment of the localized hole state is characterized by two vectors,
one in spin- and one in real space. The real space vector 
characterizing the dipole is simply the orientation of the ferromagnetic
bond in the Aharony picture while it is determined by the four
coefficients $c_{{\bf k}_0}$ and by the equivalent wave vectors
of the bound hole wave function in the quantum mechanical model. 
The coupling to the spin background
is then identical in both models.
Here we simply assume that the
localized hole produces dipolar frustration
and, rather than relating our phenomenological coupling
parameters to
a microscopic model, we derive our parameters from a comparison to
experiments.
As we discuss
below, the dipole model can quite well explain all the important
characteristics of the magnetism of the weakly doped AF and SG phase.
Let us further mention that
for Sr doping, it was proposed that a chiral spin current is induced on the
four Cu sites closest to the Sr impurity which leads to
 a Skyrmion-like distortion of the AF,
where the mechanism of frustration is again the coupling between
spin and background magnetization currents.\cite{gooding91} 

In section \ref{afphase} we introduce the dipolar frustration model,
summarize the main results of previous studies on this model, and
discuss how they compare with experiments. In section \ref{noncollinear}
we first derive an extension of the model to allow for
non-collinear correlations which arise from dipole ordering.  
We perform a RG calculation to understand the influence of disorder 
and discuss the importance of topological defects of the spin texture.
Finally, in section \ref{compexp} 
our results are compared with neutron scattering data on the SG phase
of $\rm La_{2-x}Sr_x CuO_4$. We find that the SG phase can be
described as a strongly disordered spiral phase in which topological
defects proliferate.

\section{The AF phase and dipolar frustration 
models}
\label{afphase}
We briefly sketch here the basis of the dipolar frustration model
and the results of previous studies of this model in the collinear limit.
The model as presented in this section is applicable only for the 
antiferromagnetic phase in which the dipoles do not have a preferred
direction. At high temperatures, the collinear theory
can then be used.
We will show in the next section however, that the collinear
model is not able to describe the low temperature and/or
strong disorder regime, where non-collinear behavior emerges.

In the dipole model, it is assumed that each localized hole produces
dipolar frustration. It is then possible to study the magnetism
of the hole doped materials completely ignoring the charge degrees
of freedom and to work with the spin sector only. Further, 
as there are clear indications of static AF correlations for $x<0.05$,
the antiferromagnet should be well described within the renormalized
classical regime of the QNL$\sigma$M. In this regime, 
quantum fluctuations simply lead to a renormalization of the
coupling constant of the classical model. A classical model should
thus suffice to describe the relevant physics in the AF and SG
regime.

\subsection{Ferromagnetic bonds as an example of dipolar frustration}
\label{fbonds}
\begin{figure}[htb]
\begin{center}
\includegraphics[width=5cm]{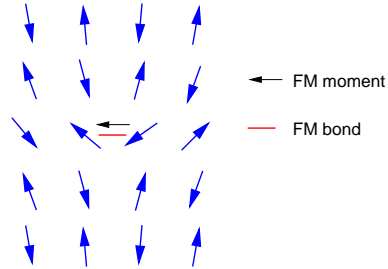}
\end{center}
\caption[]{\label{dipolst} Dipolar distortions produced by
a ferromagnetic bond}
\end{figure}

Dipolar frustration was first discussed in the general
context of insulating spin glasses by Villain.\cite{villain79}
The simplest way of producing dipolar spin textures is by
placing a ferromagnetic bond in an otherwise AF magnet, whose
order parameter we denote by $\bf n$. At a distance $\bf x$ away
from the ferromagnetic bond,
this leads to a deviation of the N\'eel order 
$\delta {\bf n}\sim {\bf f}_\mu x_\mu/x^2$. Here, ${\bf f}_\mu$
is a vector both in spin and lattice space, where $\mu=1$,2 are the
indices of the 2D lattice vector. The spin part
corresponds to the local ferromagnetic moment (with ${\bf f}_\mu \perp
{\bf n}$) produced by the bond
and the lattice part corresponds to the orientation of the bond
on the lattice (see Fig.~\ref{dipolst}). 
This can be easily
derived in a harmonic continuum
approximation, where the energy density
of the magnet away from the impurity
is proportional to $[\partial_\mu (\delta {\bf n})]^2$
and the classical equation of motion is $\nabla^2 (\delta {\bf n})=0$.
For any impurity distribution, the solution for $\delta {\bf n}$
can thus be written in a multipole expansion. As the monopole
moment is energetically too costly\cite{villain79}
the lowest order contributions, consistent with the symmetry of
the one-bond problem, are dipolar.

\subsection{Collinear Model}
Because of the long-range nature of dipolar frustration,
a continuum field theory, such as a (classical) non-linear $\sigma$-model
(NL$\sigma$M), should be well suited for a treatment of this problem.
While the dipole spin structure discussed above 
is a solution of the harmonic theory, it is
not a solution  of the 2D NL$\sigma$M.
Nonetheless one can study the dipole model within the
NL$\sigma$M, if one introduces the dipolar frustration through
a minimal coupling scheme.
As mentioned in Ref.\ [\cite{glazman90}], the dipolar frustration can be
reproduced (on the harmonic level)
via a coupling of the dipoles to the gradient of the
order parameter ${\bf n}$ of the NL$\sigma$M. Thus, within a
NL$\sigma$M approach, the reduced Hamiltonian of the model
can be written as \cite{glazman90,cherepanov99} (the factor
$\beta=T^{-1}$ is included in the Hamiltonian and we set $k_B$=1)
\begin{equation}
\label{dipham}
H_{\rm col} 
=\frac{\rho_s}{2T} \int d^2{\bf x} \left( \partial_\mu {\bf n} \right)^2
+\frac{\rho_s}{T}\int d^2{\bf x} \ {\bf f}_\mu \cdot \partial_\mu {\bf n} 
\end{equation}
where ${\bf n}^2=1$, $\rho_s$ is the spin stiffness (renormalized by quantum fluctuations),
$T$ the temperature,
$\bf n$ a three component unit vector representing the local
staggered moment 
and ${\bf f}_\mu$ is a
field representing the dipoles. We did not include here small
corrections which lower the spin symmetry from Heisenberg
to XY or Ising. While these are known to be present both in the undoped
and weakly doped compounds \cite{ando01}, they have a very small
characteristic energy scale and, as a first approximation, we set them
to zero. Note however that these terms dominate the static magnetic
susceptibility near the N\'eel transition.
For a random distribution
of localized dipoles we write
\begin{eqnarray}
{\bf f}_\mu({\bf x})={\cal M} \sum_i \delta({\bf x}- {\bf x}_i)
a_\mu({\bf x}_i) {\bf M}_i
\end{eqnarray}
where the sum is over the impurity sites,
${\bf a}_i$ are lattice unit vectors, ${\bf M}_i$ unit vectors in
spin space, and ${\cal M}$ measures the strength of the dipoles.
While there is no dipole-dipole interaction term in Eq.~(\ref{dipham}),
fluctuations of the $\bf n$ field generate a spin wave
mediated interaction. This can be seen once short scale fluctuations
are integrated out under a renormalization procedure.\cite{glazman90}
An integration
over the short scale fluctuations up to a scale $L\gg 1/\sqrt{x}$
(but $L\ll \xi$ where $\xi$ is the 2D spin correlation length)
leads to an effective interaction term of the form
\begin{eqnarray}
H[\{{\bf M}_i \}]=\frac{\rho_s {\cal M}^2}{2 T} \sum_{i,j} J_{ij} {\bf M}_i \cdot
{\bf M}_j
\end{eqnarray}
with 
\begin{equation}
J_{ij}= \frac{1}{2 \pi x_{ij}^2}\left( 
2 \frac{\left({\bf x}_{ij}\cdot {\bf a}_i \right)
\ \ 
\left({\bf x}_{ij}\cdot {\bf a}_j \right)}{x_{ij}^2}
- {\bf a}_i \cdot {\bf a}_j \right),
\end{equation}
and ${\bf x}_{ij}={\bf x}_i - {\bf x}_j.$
Thus, for an average separation of dipoles $\sim 1/\sqrt{x}$
there is a random interaction among dipoles with a characteristic
energy $U\sim \rho_s {\cal M}^2 x /{4\pi}$. It was further shown
 \cite{glazman90}
 that at high temperatures, $U\ll T$, the presence
of dipoles lead to a renormalized effective stiffness $\rho_{\rm eff}
= \rho_s (1-U/T)$. Thus, the correlation length at high temperatures
(and small $x$) has the form
\begin{equation}
\label{corrlength}
\xi\sim \exp \left( \frac{2 \pi \rho_{\rm eff}}{T} \right)
=\exp \left( \frac{2 \pi \rho_{s}}{T}-\frac{2 \pi \rho_{s} U }{T^2}
\right).
\end{equation}
This result agrees to lowest order in $x$ with that obtained 
by Cherepanov {\em et al.} 
\cite{cherepanov99} in a related renormalization group (RG) calculation
where they calculated $\rho_{\rm eff}$ up to order  $x^2$.
>From a comparison with correlation lengths obtained from
neutron scattering data at high
temperatures, they estimated $U\sim 20 \rho_s x$. The doping
dependence of $T_N$ was also found to be consistent with
the dipole model.\cite{cherepanov99}

\begin{figure}[htb]
\begin{center}
\includegraphics[width=5cm]{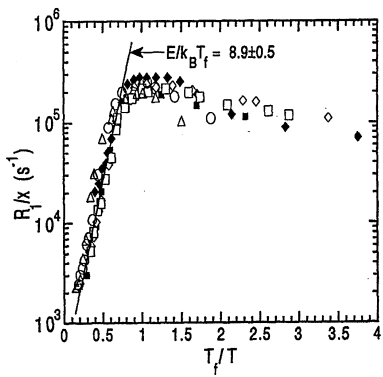}
\end{center}
\caption[]{\label{relax}
$R_1=(T_1^*)^{-1}$ data from \raisebox{1ex}{\tiny 139}La NQR relaxation
measurements for $\rm La_{2-x}Sr_x CuO_4$ and various $x<0.02$, from 
Ref.\ [\cite{chou93}].}
\end{figure}

A second independent test of the value of $U$
is to consider the spin relaxation times inside the AF phase.
This can be understood already within the
theoretical framework just presented, 
using arguments similar to those from Ref.~\cite{gooding94} where spin
relaxation has been discussed within a slightly different
frustration model.
The relaxation rates inside
the AF phase can be explained within the dipole theory if
one assumes that the relaxation is driven by the interaction
among dipoles and hence controlled by the parameter $U$. At temperatures
well above the actual freezing temperature, an Arrhenius law
is observed, with a characteristic energy $E= 8.9 T_f \sim
7250$K$x$,
see Fig.~\ref{relax}.
The above estimate of $U$ correctly reproduces 
the linear scaling of the relaxation energy with $x$ and
also gives a good estimate for the slope. 
With $U=20 \rho_s x$,
$\rho_s\sim 24$ meV \cite{cherepanov99}
one obtains $U \sim 5500$K$x$.
Considering that this is a very rough approximation, the value 
is not too far off from the experimental one.  
We mention further that the linear
scaling of the width of the distribution of local staggered
moments is also consistent with a dipole model.\cite{goldenberg97}

\section{Non-collinear correlations and dipole ordering}
\label{noncollinear}

While the dipole model presented above can well explain
the temperature and doping dependence of the correlation
length not just in the AF but also, to some extent, 
in the SG regime,\cite{cherepanov99}
theoretical investigations of the model have always
predicted (or rather assumed) short ranged commensurate antiferromagnetism.
The recent observation of incommensurate (IC) correlations
for the regime $0.02<x<0.05$ requires therefore a 
new approach to the SG phase.\cite{hasselmann00a} 

As a possible explanation for the presence of IC
correlations, a disordered striped phase has been proposed,
similar to the ordered striped phase found near $x\sim 1/8$.
While there is indeed an instability in the striped phase
toward a disordered phase at low $x$,\cite{hasselmann99} 
it is unlikely that the stripes will survive in presence 
of strong disorder. In fact, recent numerical simulations of 
Shraimann-Siggia dipoles with disorder 
have shown that the latter leads to a destruction of the stripe 
phase.\cite{branko} 

In the spin glass regime, there are two competing length scales.
The first is related to the average separation between
disorder centers (Sr ions) $\ell_d$ which scales as 
$\ell_d \sim 1/\sqrt{x}$. The other
is the scaling of the periodicity $\ell_s$ 
associated with the incommensurability, 
which scales as $\ell_s \sim 1/x$. For small $x$, $\ell_d \ll \ell_s$.
In a stripe scenario the charge distribution would also have
a periodicity which scales with $\ell_s$. Thus,
in a striped phase the charge can not take full advantage of
the disorder. The stripes must either break up into short
segments or reduce their on-stripe charge density considerably
to take advantage of the disorder potential. 
Instead we propose here a theory in which the charges are
completely disordered and the incommensurability exists
only in the spin sector. Then, there is no conflict between
the two scales $\ell_s$ and $\ell_d$ as $\ell_s$ relates only
to the spins whereas $\ell_d$ characterizes the charge distribution.

Notice that even in the case that short segments of
stripes should be present, these stripes would lose their 
anti-phase domain wall
character and instead act like a row of ferromagnetic bonds,
again causing dipolar frustration. Thus, the theory we present
here applies both to the case of localized hole states which
produce dipolar frustration as it does to a system of randomly
placed stripe segments. We view the scenario of localized holes
however as the more plausible one.

\subsection{Dipole ordering}
It is easy to see how the dipole model can lead to 
IC correlations.\cite{shraiman89a} The Hamiltonian Eq.~(\ref{dipham}) favors
the formation of a spiral phase, with a
non-zero average twist $\partial_\mu {\bf n}$ of the AF order and a
simultaneous alignment of the dipoles, 
$\left< {\bf f}_\mu \right> \neq 0$, as long as
the lattice and spin degrees of freedom of dipoles are annealed and
free to orient themselves.
 The lattice position of the $\rm Sr$ dopants 
(located above the center of a $\rm Cu$ plaquette), which 
pin the holes, suggests that this freedom indeed exists. We emphasize 
that a spatially homogeneous distribution of dipoles is not required
for the formation of spiral correlations.

The preferred orientation of the lattice part of the ${\bf f}_\mu$ vector 
is determined by the nature of the localized hole state and therefore
should reflect
the symmetries of the underlying lattice. Thus a discrete set of
favored lattice vectors for the formation of the spiral exists. 
The $a$-$b$ (or  square lattice) symmetry breaking associated with the
formation of spiral correlations can therefore have truly long
range order. 
The continuous symmetry of spin space
on the other hand
inhibits long-range magnetic order in the 2D system
for either finite temperatures or disorder. The 
experimental observation of a macroscopic $a$-$b$ asymmetry 
\cite{matsuda} but very short
spin correlation lengths thus clearly motivate the study of the dipole model.

\subsection{Continuum description of spiral phases}

We here investigate the dipole model allowing for the 
presence of non-zero ordered
moments but assume a random spatial distribution of the dipoles. 
First, however, we need  a proper theoretical
description of the homogeneous spiral phase.

\begin{figure}[htb]
\begin{center}
\includegraphics[width=5.5cm]{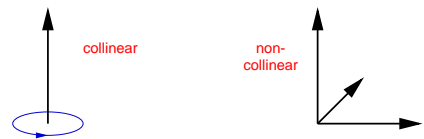}
\end{center}
\caption[]{\label{noncolpic}The order parameter of collinear magnets, which
are invariant under rotations around the collinear axis, can be 
represented by a unit vector (left), whereas non-collinear order parameters
require three orthonormal vectors (right).}
\end{figure}

In collinear magnets, the rotational $O(3)$ symmetry of the 
system is broken down to a  ground state with $O(2)$ symmetry, as rotations
around the magnetization axis leave the ground state invariant
(this is schematically shown on the left hand side of Fig.~\ref{noncolpic}).
The order parameter of collinear magnets is then an element
of $O(3)/O(2)$. This group is isomorphic to the group of three dimensional
unit vectors $\bf n$, which is the representation used in the Hamiltonian
Eq.~(\ref{dipham}). Further, in absence of dipoles, the Hamiltonian
Eq.~(\ref{dipham}) is invariant with respect to $O(2)$ rotations of the
lattice variables.  The spin and lattice symmetries are decoupled and
independent for the collinear AF. 
A spiral ground state on the other hand breaks the
$O(3)$ spin symmetry completely. Moreover, in a spiral state  
the lattice symmetries and the spin symmetries are no longer decoupled
and the order parameter space of such a state becomes more involved. 

For spirals, 
the combined symmetry of lattice and the spin space is $O(3)\times O(2)$.
As discussed in detail by Azaria {\em et al.},\cite{azaria93} 
the coupling of the
spin and lattice degrees of freedom in frustrated spin systems leads
to an order parameter which results from a symmetry breaking
of the combined lattice and spin degrees of freedom and is in 
general of the form 
$O(3)\times O(q)/O(q)$ where $q$ depends on the symmetries
of the lattice. 
For a  spiral phase, one finds \cite{klee96} $q=2$.

\begin{figure}[htb]
\begin{center}
\includegraphics[width=7cm]{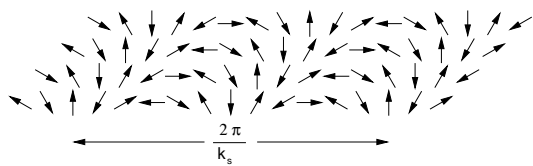}
\end{center}
\caption[]{\label{spiralpic} Spin texture of an AF spiral.}
\end{figure}

A convenient representation of the order
parameter is in terms of orthonormal ${\bf n}_k$, $k=1\dots3$, with
$n^a_k n^a_q=\delta_{kq}$. Klee and Muramatsu  \cite{klee96}
have derived a continuum field theory for the ${\bf n}_k$
order parameters from the lattice Heisenberg model Eq.~(\ref{sqlh}), assuming
an IC spiral state with an ordering wave vector
${\bf k}_S=(\frac{\pi}{a},\frac{\pi}{a})+
{\bf q}_S$. Here, ${\bf q}_S$ measures
the deviation from the commensurate AF wave vector, see 
Fig.\ \ref{spiralpic}. The 
lattice spins ${\bf S}_i$ at sites ${\bf r}_i$ can be parametrized
in a spiral configuration with the use of the $n^a_k$ as (with 
${\bf n}_3={\bf n}_1 \times {\bf n}_2$)
\begin{eqnarray}
\label{klee0} 
{\bf S}_{i}/S= {\bf n}_1 \cos ({\bf k}_S \cdot {\bf r}_{i})
- {\bf n}_2 \sin ({\bf k}_S \cdot {\bf r}_{i}).
\end{eqnarray}

A perfectly ordered spiral is described by Eq.~(\ref{klee0}) with
constant, i.e. space independent, ${\bf n}_k$. To allow for
spatial fluctuations of the spins around the spiral order, 
Klee and Muramatsu introduced 
a slowly varying field ${\bf L}$ via \cite{klee96,dombre89}
\begin{eqnarray} \nonumber
\label{klee1}
\frac{{\bf S}({\bf r}_i)}{S}
&=&\frac{{\bf N} + a {\bf L}}
{\sqrt{1 + 2 a {\bf N}\cdot {\bf L} + a^2 L^2}}
\\ \nonumber 
&=& {\bf N} + a \left[{\bf L} - \left({\bf N} \cdot {\bf L}\right) \right]
-a^2 \left[ \left({\bf N}\cdot {\bf L}\right) {\bf L}
+\frac{1}{2} {\bf L}^2 {\bf N} \right. \\ &-& \left. 
\frac{3}{2} \left({\bf N}\cdot {\bf
L}\right)^2 {\bf N} \right] + {\cal O}(a^3),
\end{eqnarray}
where ${\bf N}={\bf n}_1 \cos ({\bf k}_S \cdot {\bf r}_{i})
- {\bf n}_2 \sin ({\bf k}_S \cdot {\bf r}_{i})$ with
now slowly fluctuating fields ${\bf n}_k$.
The continuum theory
can then be found upon expressing in the lattice
Heisenberg model the spin operators
in terms of the ${\bf n}_k$  and $\bf L$ fields,
expanding the terms up to order $a^2$ and taking the limit $a\to 0$
in the end. After integrating out the ${\bf L}$ fields,
one finds an effective Hamiltonian which can be 
written in the classical limit in the general form \cite{klee96}
(again we include the factor $\beta=T^{-1}$ into $H$)
\begin{equation}
\label{spiral1}
H=
\frac{1}{2} \int d^{2}{\bf x} \ p_{k\mu} (\partial_{\mu} {\bf n}_{k})^{2}
+ s_\mu \int d^2{\bf x} \  {\bf n}_1 \cdot \partial_\mu {\bf n}_2.
\end{equation}
This description is valid for length scales larger than
$|{\bf q}_S|^{-1}$. 
The stiffnesses of the order parameter ${\bf n}_k$ are given
initially by $p_{1\mu}=p_{2\mu}=J S^2 \cos(q_{S\mu}a) /(2T)$ and
$p_{3\mu}\simeq0$, but will change under a renormalization of the
model. We will ignore for the most part the 
small anisotropy (of order $q_S^2a^2$)
in the stiffnesses $p_{k\mu}$ and just write $p_k$. 
The vector $\bf s$ is to lowest order
given by ${\bf s}= J {\bf q}_S/T$. 
The term with the $s_\mu$ pre-factor
makes this Hamiltonian unstable, which simply expresses the fact
that the pure Heisenberg model does not support a
spiral phase ground state.
The $s_\mu$ term will however be
canceled by a similar term originating from the coupling of
the spins to the ordered fraction of the dipoles, relating the 
incommensurability self-consistently to the ordered moment of the
dipoles. In other words, the ordered dipoles stabilize the spiral
phase, as expected. 

It must be stressed that because the continuum model is only valid
at length scales larger than the period of the IC structure,
there is a relatively large uncertainty in the estimates of the $p_{k\mu}$.
There is always a fundamental
problem in relating the continuum model parameters to those of 
the original microscopic lattice
model, but in this case this problem is especially severe.
The continuum model parameters must be obtained from an
average over one period of the spiral which,
for small incommensurabilities, can be rather large.
Thus,
the above estimates for the $p_{k\mu}$'s should be taken with care.

\subsection{Disorder coupling: a gauge glass model}
We now must include the coupling of the dipolar frustration centers
to the spiral order parameter. While there is no microscopic derivation
of this coupling at hand, the fact that the coupling in the collinear
model can be expressed within a minimum coupling scheme allows for
a simple generalization of the model to non-collinear spin states.
We first observe that the ordering wave vector of the spiral, ${\bf q}_S$,
is entirely determined by the average orientation of the dipoles.
Similarly,
local variations of the density or orientation of the
dipoles should also modify the local  ordering
wave vector. Further, to reproduce the strong canting produced
by the dipoles, the coupling should be of first order in the spatial
derivative of the spiral order parameter. 
To generate the frustration produced by the dipoles we thus
introduce a minimal coupling \cite{hertz78} in the first term
of Eq.~(\ref{spiral1}), i.e. we replace $(\partial_\mu {\bf n}_k)^2$
with
$[ (\partial_\mu - i {\bf B}_\mu \cdot {\bf L} ){\bf n}_k]^2$
where ${\bf B}_\mu$ is a random gauge field, representing the dipoles. 
The components of $\bf L$ are $3\times 3$
matrix representations of angular momenta which generate rotations about
the three spin axes, with
\begin{eqnarray}
-i {\bf B}_\mu \cdot {\bf L} \  {\bf n}_k=
{\bf B}_\mu \times {\bf n}_k.
\end{eqnarray}
This coupling has the advantage of relative
simplicity combined with a clear physical interpretation: the dipolar
fields define the locally preferred wave vector of the spiral, and
fluctuations of the dipole fields lead to fluctuations of the wave vector.
Further, it reproduces the correct form of the dipole coupling in the
collinear limit, as shown below.
Let us write 
${\bf B}_\mu = \left[{\bf B}_\mu \right]_D + {\bf Q}_\mu$
so that $\left[{\bf Q}_\mu \right]_D=0$, where $\left[ \dots \right]_D$ 
is the disorder average. 
We then obtain the following Hamiltonian for the spiral
in presence of disorder,
\begin{equation}
\label{spiral2}
H=
\frac{1}{2} \int d^{2}{\bf x}  p_{k\mu} (\partial_{\mu} {\bf n}_{k})^{2}
+ \int d^{2} {\bf x} \ p_{k}  \partial_{\mu} {\bf n}_{k} \cdot {\bf Q}_{\mu}
\times {\bf n}_{k} ,
\end{equation}
where the ordered part of the dipole field
cancels the second term in Eq.~(\ref{spiral1}). Thus,
\begin{eqnarray}
p_{k\mu} \partial_\mu {\bf n}_k \cdot 
\left[{\bf B}_{\mu} \right]_D \times {\bf n}_k + 
s_\mu {\bf n}_1 \cdot \partial_\mu {\bf n}_2=0.
\end{eqnarray} 
As ${\bf q}_S \propto {\bf s}$, this equation relates the incommensurability
linearly to the density of ordered dipoles.
The remaining part of the dipole field, $\bf{Q}_\mu$, is a quenched
variable with zero mean and we assume Gaussian short ranged statistics,
\begin{eqnarray}
\label{defdis}
\left[Q_{\mu}^{a}({\bf x}) Q_{\nu}^{b}({\bf y)}\right]_D=\lambda
 \delta({\bf x-y})\ \delta_{ab}\ \delta_{\mu \nu}.
\end{eqnarray}
In absence 
of disorder, the Hamiltonian defined by Eq.~(\ref{spiral2}) has the desired
$O(3)\times O(2)/O(2)$ symmetry.
The $O(3)$ symmetry is associated with the spin indices $a$ of the
$n_k^a$, while
the $O(2)$ symmetry is associated with the lattice indices $k$ and
arises because $p_{1\mu}=p_{2\mu}$.
Hence, the equality $p_{1\mu}=p_{2\mu}$ is  directly related and enforced
by the symmetries of the spiral. Note that if all $p_{k\mu}$ are
identical, the lattice symmetry is enhanced to $O(3)$.
We further see now, that 
the model reduces to the collinear model 
Eq.~(\ref{dipham}) in the case $p_{1,2}=0$ with
$p_3=\rho_s/T$, ${\bf n}_3={\bf n}$ and ${\bf f}_{\mu} =
{\bf Q}_{\mu} \times {\bf n}$. Unfortunately it is
not possible to reach
the collinear limit by sending ${\bf q}_S\to 0$. The reason is that
the parameters $p_{k\mu}$ are, within the approximation used in their
derivation, independent of the size of the unit cell of the spiral,
i.e. in the limit ${\bf q}_S\to 0$, the unit cell size diverges
while the parameters $p_{k\mu}$ remain unaffected. 

The model defined by Eq.~(\ref{spiral2}) is in fact far more general
than its derivation might suggest. In absence of disorder it is
applicable to other types of frustrated spin systems with a 
non-collinear ground-state, such as e.~g. the Heisenberg model on
a triangular lattice.\cite{dombre89,apel92,azaria93} It is conceivable,
that certain types of randomness in such lattices may be well described
by the disorder coupling employed  here. More importantly, the
model Eq.~(\ref{spiral2}) can be viewed as a general model to
investigate diluted spin glasses, in which a spin system is
frustrated by a small number of impurities. There have been 
investigations of similar models of spin glasses in the past,
most notably by Hertz,\cite{hertz78} which however did not account for 
non-collinear correlations which are 
known to be essential
in spin glasses.\cite{binder86}
Our approach has the appeal that it can interpolate between collinear
and non-collinear states and thus offers the possibility to study
the transition from an ordered collinear magnet to a disordered 
non-collinear one continuously.

\subsection{Renormalization}
We now investigate the renormalization of the model under
a change of scale, with the objective to understand the influence
of the dipoles on the correlation length of the model. 
For carrying out the RG calculation,  it is of advantage to use a
$SU(2)$ representation of the model \cite{apel92}
(see also App.~\ref{appsu2}). We therefore 
write  
\begin{eqnarray}
\label{su2n}
n_k^a =  \frac{1}{2}\mbox{tr} \left[ \sigma^a  
g  \sigma^k g^{-1} \right]
\end{eqnarray}
where $\sigma^a$ are Pauli matrices and $g\in SU(2)$. We further introduce
the fields \cite{polyakov75}
\begin{eqnarray}
\label{su2a}
A_\mu^a = \frac{1}{2i} \mbox{tr} \left[ \sigma^a  g^{-1}  
\partial_\mu g \right], 
\end{eqnarray}
which are related to 
the first spatial derivatives of ${\bf n}_k$ through 
$\partial_\mu n_k^a = 2 \epsilon_{ijk} A_\mu^i n_j^a$.
Eq.~(\ref{spiral2}) then acquires the form,
\begin{eqnarray} 
\label{finalspiral}
H&=&\frac{1}{t_\mu} \int d^{2}{\bf x} \left[ {\bf A}_\mu^2 
+ b {A_\mu^z}^2
\right] + \nonumber \\ && +
2 \int d^{2}{\bf x} \ p_{k\mu} \   \epsilon_{ijk} \ \epsilon_{abc} \
 A_\mu^i \ n_j^a \ n_k^c \ {Q}_\mu^b \ . 
\end{eqnarray}
where 
$t_\mu^{-1}=2(p_{1\mu}+p_{3\mu})$ 
and $b=(p_{1\mu}-p_{3\mu})/(p_{1\mu}+p_{3\mu})$.
At the point $b=0$ 
the symmetry is enhanced to $O(3)\times O(3)/O(3) \simeq O(4)/O(3)$
while at $b=-1$ the model is collinear. For spirals, we have initially
$b=1$.

We first discuss the dimensional scaling behavior of the model 
(\ref{spiral2}, \ref{finalspiral}). We assign the dimension $-1$ 
to each spatial
dimension so $\partial_\mu$ has dimension 1. It
follows that the ${\bf A}_\mu$ fields
have a
scaling dimension of 1. The scaling dimension of 
the
first term in Eqs.~(\ref{spiral2},~\ref{finalspiral}) 
is then $2-d$ where here $d=2$.
Thus, this term is marginal and
an RG analysis is required to study the scaling of the $t_\mu$, $b$
parameters. 
Local terms containing more than two ${\bf A_\mu}$ terms have 
positive dimensions and are irrelevant.
Hence, such
terms, while they are generated in the perturbative expansion we discuss
below, need not be
considered.

As was pointed out in Ref.\ [\cite{cherepanov99}], for the disorder choice
(\ref{defdis})  the model defined by Eq.~(\ref{dipham})
has a lower critical dimension of two and is thus renormalizable in
two dimensions, as can be shown with a general
Imry-Ma type argument. The same argument can be used for the present
model.
The disorder coupling in Eq.~(\ref{spiral2}) can 
be rewritten in momentum space as a random field coupling 
of the form 
\begin{eqnarray}
\label{rfc} &&
\int \frac{d^2{\bf q}}{(2 \pi)^2} \ \ \  
{\bf n}_k({-\bf q}) \cdot {\bf h}_k({\bf q}); \nonumber  \\
&&
{\bf h}_k({\bf q})= i p_{k\mu} q_\mu \int d^2{\bf x}\ 
({\bf Q}_\mu \times {\bf n}_k) e^{i {\bf q }\cdot {\bf x}} \nonumber
\end{eqnarray}
where
the random fields ${\bf h}_k({\bf q})$ have disorder correlations with
a momentum dependence
$\left[ {h}_k^a({\bf q}) {h}^{a^\prime}_{k^\prime}({\bf q}^\prime) \right]_D
\propto \delta({\bf q}-{\bf q}^\prime) |{\bf q}|^\Theta$ with $\Theta=2$. 
According
to general arguments by Imry and Ma,\cite{imry75}
in models with continuous symmetries
random fields will destroy 
long-range order as long as $d< 4-\Theta$. This implies that in our case
$d=2$ is the lower critical dimension \cite{cherepanov99} and a
renormalization group analysis of both the stiffness  and the disorder 
coupling is required.

We now derive the one-loop RG equations. For this, we 
split the original $SU(2)$ field $g$ 
into slow and fast modes,
$g=\tilde{g} \exp (i$ $\varphi^a   \sigma^a )$ and trace out the fast 
modes $\varphi^a$ which have fluctuations in the range 
$[\Lambda^{-1},1]$, where we set the 
original UV cutoff equal to 1.
For the one-loop calculation, we need an expansion of Eq.~(\ref{finalspiral})
up to second order in $\varphi^a$ (higher order terms will only contribute
at higher loop order of the RG).
For the fields ${\bf n}_k$ and
${\bf A}_\mu$ the expansion reads (see App.~\ref{expphi} for more
details)
\begin{eqnarray}
A_\mu^i
&=& \tilde{A}_\mu^i + \partial_\mu \varphi^i 
+\epsilon_{ijk} \varphi^j \partial_\mu \varphi^k
+2 \epsilon_{ijk}\varphi^j \tilde{A}_\mu^k
 -2 \tilde{A}_\mu^i\ \fett{\varphi}^2 \nonumber \\ \nonumber
&+& 2 \tilde{\bf A}_\mu\fett{\cdot \varphi} \ \varphi^i+ {\cal O}(\varphi^3), 
\\ 
n_i^a &=& 
\tilde{n}_i^a +2 \epsilon_{ijk} \varphi^j \tilde{n}_k^a + \varphi^j 
\varphi^k R_{jk}^{ai}
+ {\cal O}(\varphi^3), \nonumber
\end{eqnarray}
where
$$
R_{jk}^{ai}= \frac{1}{2} \ \mbox{tr} \left\{ \sigma^a \tilde{g}
\left( \sigma^j \sigma^i \sigma^k -\frac{1}{2} \sigma^j \sigma^k
\sigma^i - \frac{1}{2} \sigma^i \sigma^j \sigma^k \right)\tilde{g}^{-1} 
\right\}.
$$
The expansion of the energy functional (\ref{finalspiral}) reads
\begin{equation}
\label{pertexp}
H =\frac{1}{t_\mu} \int d^2{\bf x} 
\left[\tilde{\bf A}_\mu^2 + b \left(\tilde{A}_\mu^z\right)^2 \right] + H_{c0}
+ H_\varphi + H_p 
\end{equation}
with
\begin{eqnarray}
H_{c0}&=&2 \int d^2{\bf x} p_{k\mu} \epsilon_{ijk} \ \epsilon_{abc}\
\tilde{A}_\mu^i \tilde{n}_j^a \tilde{n}_k^c Q_\mu^b, \nonumber \\
H_p&=&H_1 + H_2 + H_3 + H_4
+ H_{c1} +H_{c2} + H_{c3} + H_{c4}. \nonumber
\end{eqnarray} 
The first two terms in the expansion of $H$ have exactly the same form as the
original functional (\ref{finalspiral}), but are now functionals of the slow
fields. $H_\varphi$ is quadratic in $\varphi$ and has the
form
\begin{eqnarray} \nonumber
H_\varphi = \frac{1}{t_\mu} \int d^2{\bf x} 
\left[ \left( \partial_\mu \fett{\varphi} \right)^2
+ b \left( \partial_\mu \varphi^z \right)^2 \right]
\end{eqnarray}
$H_{1}\dots H_4$ are generated by the first term in Eq.~(\ref{finalspiral})
and are given by
\begin{eqnarray}
\nonumber 
H_1 &=& 2 t_\mu^{-1} \int d^2{\bf x} \ \tilde{A}_\mu^i \partial_\mu \varphi^j
\varphi^k \epsilon_{ijk} \left( 1- b \delta_{iz} + 2 b \delta_{jz}
\right), \\ \nonumber
\label{pertexp2}
H_2&=& 2 t_\mu^{-1} \int d^2{\bf x} \ \partial_\mu \varphi^i \tilde{A}_\mu^i
\left(1+b \delta_{iz}\right), \\ \nonumber
\label{pertexp3}
H_3&=& 4 b t_\mu^{-1} \int d^2{\bf x} \ \epsilon_{zjk} \tilde{A}_\mu^z \varphi^j
\tilde{A}_\mu^k, \\ \nonumber
\label{pertexp4}
H_4&=& 4 b t_\mu^{-1} \int d^2{\bf x} \ \left[
\left(\epsilon_{zjk} \varphi^j \tilde{A}_\mu^k \right)^2 - \left(\tilde{A}_\mu^z\right)^2
\fett{\varphi}^2 \right. \\ \nonumber
&+& \left. \tilde{A}_\mu^z \varphi^z \tilde{\bf A}_\mu\fett{\cdot \varphi}
\right].
\end{eqnarray}
The coupling term in Eq.~(\ref{finalspiral}) produces the $H_{c1}\dots H_{c4}$
terms,
\begin{eqnarray} \nonumber
H_{c1}&=&4 \int d^2{\bf x} \ p_{k\mu}\epsilon_{ijk}\epsilon_{abc}
\left[ \epsilon_{klm} \tilde{n}_j^a \tilde{n}_m^c \tilde{A}_\mu^i  \right. \\ \nonumber &+&  \left.
\epsilon_{jlm} \tilde{n}_k^c \tilde{n}_m^a  \tilde{A}_\mu^i
+ \epsilon_{ilm}
\tilde{n}_j^a \tilde{n}_k^c  \tilde{A}_\mu^m\right] \varphi^l Q_\mu^b, 
\\ \nonumber
H_{c2}&=& 2 \int d^2{\bf x} \ p_{k\mu}\epsilon_{ijk}\epsilon_{abc}
\partial_\mu \varphi^i \tilde{n}_j^a \tilde{n}_k^c Q_\mu^b, \\ \nonumber
H_{c3}&=& 2\int d^2{\bf x} \ p_{k\mu}\epsilon_{ijk}\epsilon_{abc}
\left[ \tilde{A}_\mu^i \left( \tilde{n}_j^a R_{lm}^{ck} 
+\tilde{n}_k^c R_{lm}^{aj} \right)
\varphi^l \varphi^m \right. \\ \nonumber
&+& 
2 \tilde{n}_j^a \tilde{n}_k^c \left( \tilde{\bf A}_\mu\fett{\cdot \varphi}\
\varphi^i - \tilde{A}_\mu^i \fett{\varphi}^2 \right) 
 \\ \nonumber
&+&   4 \left(\tilde{A}_\mu^i \epsilon_{jlm} \epsilon_{kpq} 
\tilde{n}_m^a \tilde{n}_q^c 
+ \tilde{A}_\mu^m \epsilon_{ilm} \epsilon_{kpq} \tilde{n}_q^c \tilde{n}_j^a 
\right. \\ \nonumber &+&  \left. \left. 
\tilde{A}_\mu^m \epsilon_{ilm} \epsilon_{jpq} \tilde{n}_q^a \tilde{n}_k^c 
\right) \varphi^p \varphi^l \right] Q_\mu^b, \\ \nonumber
H_{c4}&=& 2\int d^2{\bf x} \ p_{k\mu}\epsilon_{ijk}\epsilon_{abc}
\left[ 2 \left( \epsilon_{klm} \tilde{n}_j^a \tilde{n}_m^c 
\right. \right.   \\ \nonumber 
&+& \left. \left. \epsilon_{jlm}\tilde{n}_m^a \tilde{n}_k^c \right)
\partial_\mu \varphi^i \varphi^l
+ \epsilon_{ilm} \partial_\mu \varphi^m
\varphi^l \tilde{n}_j^a \tilde{n}_k^c \right] Q_\mu^b.
\end{eqnarray}

The integration over the fast $\varphi$ fields is performed with
$$
\int {\cal D}[\varphi^i] \exp(-H_\varphi) \exp(-H_p) =
e^{-\cal F} \int {\cal D}[\varphi^i] 
\exp(-H_\varphi)
$$
where $\cal F$ is obtained from a cumulant expansion
\begin{eqnarray} \nonumber
-{\cal F} &=& \ln \frac{\int {\cal D}[\varphi^i] \exp(-H_\varphi) \exp(-H_p)}
{\int {\cal D}[\varphi^i] \exp(-H_\varphi)} \\ 
&=&\sum_{n=1}^\infty \frac{(-1)^n}{n!} \left< H_p^n \right>_{\varphi c}
\label{cumulant}
\end{eqnarray}
and $\left< \dots \right>_{\varphi c}$ indicates that only
connected diagrams are to be considered. 

\subsubsection{Renormalization of the spin stiffness}
\begin{figure}[htb]
\begin{center}
\includegraphics[width=8cm]{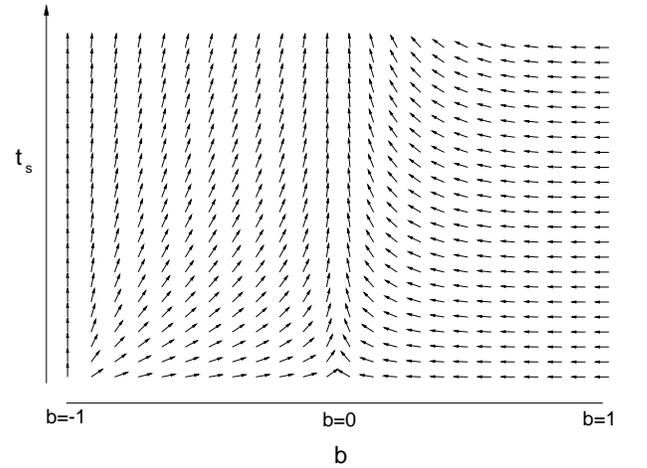}
\end{center}
\caption[]{\label{rgflow} RG flow of $t_s$ and $b$ for $\lambda=0$.
For any $b>-1$, the flow is toward $b=0$.}
\end{figure}

We ignore the (small) anisotropy of the $t_\mu$ parameter
and simply use the isotropic mean $t_s=\sqrt{t_1 t_2}$ in the
RG analysis below.
We collect all terms in the perturbative expansion which
are bilinear in $\tilde{A}_\mu^i$. After performing the
disorder average of $\cal F$, the renormalized stiffnesses
of the $\tilde{A}_\mu^i$ fields is found to be (see App.~\ref{propphi} and 
\ref{apptterms})
\begin{eqnarray} \nonumber
\frac{1}{\tilde{t_s}} &=& \frac{1}{t_s} -
\left[ \frac{2(1-b)}{t_s} + \frac{(2-b+b^2) \lambda}{t_s^2}\right] C^x({\bf 0}), \\ \nonumber
\frac{\tilde{b}}{\tilde{t_s}} &=& \frac{b}{t_s} - 
\left[ \frac{2 b (3+b)}{t_s} +
\frac{b (5+ b) \lambda}{t_s^2}  
\right] C^x({\bf 0}).
\end{eqnarray}
With $\ell = \ln \Lambda$ and 
$$
C^x({\bf 0})=\frac{t_s}{4 \pi} \ln \Lambda
$$
one then finds the RG equations
\begin{eqnarray} \nonumber
\frac{\partial}{\partial \ell} \frac{1}{t_s}&=&
-\frac{1-b}{2 \pi} -\frac{(2-b+b^2) \lambda}{4 \pi t_s}, \\ \nonumber
\frac{\partial}{\partial \ell} \frac{b}{t_s}&=&
- \frac{(3+b)b}{2 \pi}  -\frac{(5+b) b \lambda}{4 \pi t_s}.
\end{eqnarray}
This yields
\begin{eqnarray}
\label{RGT}
\frac{\partial}{\partial \ell} t_s & = & \frac{1-b}{2 \pi} t_s^2
+ \frac{2 -b + b^2}{4 \pi} \lambda t_s, \\ 
\label{RGb}
\frac{\partial}{\partial \ell} b  & = & -\frac{b (1+b)}{\pi} t_s
- \frac{b (1+b)(3-b)}{4 \pi} \lambda.
\end{eqnarray}
For $\lambda=0$, these equations describe the RG of a clean 
spiral,\cite{apel92} while for the collinear point $b=-1$, the equations
reproduce the RG of the stiffness for disordered
collinear models.\cite{cherepanov99} From Eq.~(\ref{RGb}) it is seen
that there are two fixed points for $b$
(the asymptotic freedom of the model
prevents a true fixed point in 2D as $t_s$ always
diverges).
 The collinear point $b=-1$ is 
unstable whereas $b=0$ is stable, 
irrespective of the disorder. The RG flow of $t_s$ and $b$ is shown
in Fig.~\ref{rgflow} for $\lambda=0$.  
The flow does not change qualitatively for finite
$\lambda$ as long as $\lambda\ll t_s$. Hence,     
the coupling
to weak disorder does not lead to any new fixed points, although
the disorder renormalizes the stiffness. 

\subsubsection{Renormalization of disorder coupling}
As we discuss below, the renormalization of $\lambda$ is given
by terms proportional to $\lambda t_s$ and $\lambda^2$. As the
disorder enters the renormalization of $t_s$ only in the
combination $\lambda t_s$ (see Eq.~(\ref{RGT})), 
we can neglect the renormalization
of $\lambda$ altogether for $t_s \gg \lambda$, i.~e. at high
temperatures (we have $t_s\propto T/J$).
However, for low temperatures
the renormalization of $\lambda$ must be taken into account.  
To calculate the renormalization of the disorder we follow the approach 
used in Ref.\ [\cite{cherepanov99}]. 
In this approach, the renormalized
disorder variance is defined by the variance of all terms in the
perturbative expansion
which couple
to the quenched disorder fields and are linear 
in $\tilde{\bf A}_\mu$.  
Note however that there exists no symmetry argument which guarantees
that the functional form of the disorder coupling remains unchanged
under the RG. It is thus possible that new disorder terms are
generated so that a simple renormalization of $\lambda$ is not sufficient.
This is indeed the situation we encounter for general $b\neq 0$ and discuss in
more detail below, where we find the generation of new coupling terms
at order $\lambda^2$. 
To find the complete renormalization of the
model one would have to include all generated new terms into the
original model, which is a rather laborious process which we did
not pursue. 
Nonetheless, as we have just shown above, there are only two
possible fixed points even in absence of disorder, $b=0$ and $b=-1$.
Rather than trying to categorize all possible disorder couplings, we 
therefore focus on a discussion of the RG of the disorder
near these two possible 
fixed points and discuss their stability under the flow.

We begin with the collinear case, $b=-1$. In this case,
the renormalized variance of the terms linear in $\tilde{A}_\mu^i$
is given by (see App.~\ref{applterms} and
App.~\ref{applr2}, Eq.~(\ref{rendis}))
\begin{eqnarray} \nonumber 
\frac{\lambda}{t_s^2} &\int& d^2{\bf x} \left\{ \left[
\left(\tilde{A}_\mu^x \right)^2 + \left(\tilde{A}_\mu^y \right)^2
\right] \left( 1 - \frac{2}{\pi} t_s \ln \Lambda  \right. \right. \\ &-&
\left. \left.  \frac{1}{2 \pi} \lambda \ln \Lambda \right) 
+ \left(\tilde{A}_\mu^z \right)^2 \frac{1}{2 \pi} \lambda \ln \Lambda
\right\}.
\label{rendiscol}
\end{eqnarray} 
What is evident from this result is that the renormalized
disorder coupling is no longer of the original form
$p_k \partial_\mu {\bf n}_k \cdot {\bf Q}_\mu \times {\bf n}_k$.
Such a coupling has a variance which includes a prefactor of $(1+b)^2$
of $\left(\tilde{A}_\mu^z \right)^2$.
According to Eq.~(\ref{RGb}), $b=-1$ is not changed under the influence of
the original disorder coupling. A renormalization which
retains the form of the original coupling can then not lead
to a renormalized disorder variance with a finite prefactor
of $\left(\tilde{A}_\mu^z \right)^2$ at $b=-1$. 
Such a term is however present
in Eq.~(\ref{rendiscol}) we conclude that a new
type of disorder coupling is generated at $b=-1$. 
This is perhaps easier to see in Fourier space, where the
original disorder coupling can be written as a correlated random
field coupling 
${\bf n}_k(-{\bf q})\cdot {\bf h}_k({\bf q})$, see Eq.~(\ref{rfc}).
For the original minimal coupling one has ${\bf h}_k({\bf q})\propto p_k$
and thus, in the collinear limit $b=-1$ (or $p_1=0$), only ${\bf n}_3$
is affected by this coupling. We can then interpret the finite prefactor
of the $\left(A_\mu^z\right)^2$ term in the disorder variance as the
generation of correlated fields which couple also to ${\bf n}_{1,2}$ even
at $b=-1$. It is evident that such a coupling will drive the system away
from $b=-1$ and thus destroy the collinear fixed point.  
Thus, even if the original AF order is collinear (i.e. in absence
of dipole ordering), the disorder
drives the system to a non-collinear state.
An analysis which pre-supposes 
collinear order is thus not valid in the presence of dipoles and cannot
describe the low temperature regime correctly.
Physically, one would also expect the appearance of non-collinearity.
The random 
canting of spins leads to a random local deviation of the spins from
the ordering axis and thus
destroys the remaining $O(2)$
spin symmetry of the collinear model.

To make contact with the RG result obtained from the collinear
model in Ref.\ [\cite{cherepanov99}], we note that we
can reproduce the result Cherepanov {\em et al.} obtained for the disorder
renormalization if we ignore non-collinear modes. 
We can then define the renormalization
of $\lambda$ just by the terms which are present in a purely collinear theory,
i.e. by the
$\left[
\left(\tilde{A}_\mu^x \right)^2 + \left(\tilde{A}_\mu^y \right)^2
\right]$ term in Eq.~(\ref{rendiscol}). Then
\begin{eqnarray}
\frac{\partial}{\partial \ell} \frac{\lambda}{t_s^2}&=&
- \frac{2 \lambda}{\pi t_s} -\frac{\lambda^2}{2 \pi t_s^2},
\end{eqnarray}
which, using Eq.~(\ref{RGT}) leads to
\begin{eqnarray}
\frac{\partial}{\partial \ell} \lambda &=& \frac{3}{2 \pi} \lambda^2.
\end{eqnarray}
This, together with Eq.~(\ref{RGT}) are the RG equations found in 
Ref.\ [\cite{cherepanov99}] (note that our stiffness $t_s$ differs
from the stiffness $t$ used in Ref.\ [\cite{cherepanov99}] by a 
factor two). We emphasize that this result 
ignores the role of non-collinearity in the problem. 

We now turn to the point $b=0$, the only remaining possible fixed
point of the model. 
At this highest symmetry point we find that   
no new coupling terms are generated. The variance of the
renormalized disorder coupling takes the form
\begin{eqnarray}
\frac{\lambda}{t_s^2} \int d^2{\bf x} \left\{ 
\tilde{\bf A}_\mu^2 
\left( 1 - \frac{4 t_s +3 \lambda}{4 \pi} \ln \Lambda \right)
\right\} .
\end{eqnarray}
Thus,
\begin{eqnarray}
\frac{\partial}{\partial \ell} \frac{\lambda}{t_s^2}&=&
-\frac{1}{\pi} \frac{\lambda}{t_s}  -\frac{3}{4 \pi} \frac{\lambda^2}{t_s^2}
\end{eqnarray}
which yields
the RG equation, valid for
$b=0$ but any initial ratio of $\lambda/t_s$,
\begin{eqnarray}
\frac{\partial}{\partial \ell} \lambda  &=& \frac{\lambda^2}{4 \pi}.
\end{eqnarray}
Using Eq.~(\ref{RGT}), we can simplify this
through $z=t_s+\lambda/2$ to get
\begin{eqnarray}
\frac{\partial}{\partial \ell} z = \frac{1}{2 \pi}z^2.
\end{eqnarray} 
So for $b=0$ the presence of
disorder leads to an additive renormalization of the stiffness,
$t_s\to t_s+\lambda/2$. 
In presence of any amount of disorder, 
the IC correlation length $\xi$ at $T=0$ is 
finite, as can be inferred from an integration of the RG equation with
$b=0$,
yielding
$\xi \propto \exp(C \left(t_{s0} + \lambda_0/2\right)^{-1})$  
with some cutoff dependent constant $C$.
Thus, even at $T=0$,  $\xi \propto \exp(2 C   \lambda_0^{-1})$ 
is finite.
While the disorder scales to strong coupling,
the relative disorder strength with respect
to the stiffness, $\lambda/t_s$, always scales to zero so that at
long wavelengths the disorder becomes less relevant. This is surprisingly
different to the situation with $b=-1$ fixed,\cite{cherepanov99}
where the ratio $\lambda/t_s$ was found to diverge below a certain
initial value of $\lambda_0/t_{s0}$ which was interpreted as the scaling
toward a new disorder dominated regime. Thus, if one correctly takes into
account the non-collinearity, this disorder dominated phase disappears.
The absence of 
a sharp cross over from a weak disorder to a strong disorder regime
is certainly surprising, especially as the experiments clearly observe
a transition into a spin glass phase at a finite 
temperature.\cite{wakimoto} The finite temperature transition may be related
to the presence of inter-layer coupling. We argue below, however, that
topological defects can alter the RG behavior considerably and may be
a more natural explanation for the appearance of a strong disorder regime.

\subsection{Topological defects: saddle point treatment}
The RG results presented above do not take into account topological
defects \cite{wintel94}
of the spiral as only spin waves excitations enter the calculation.
As is well known from XY spin models,  topological defects
can play an important role and drive finite temperature 
transitions.\cite{kosterlitz73}
The neglect of topological defects has been a source of 
criticism toward the 
NL$\sigma$M approach to frustrated  magnets, 
which gives controversial results for $\epsilon=1,2$ in
an $\epsilon$ expansion around $D=2+\epsilon$ 
dimensions.\cite{kawamura98} 
For two dimensional systems, the 
NL$\sigma$M results were however found to be in very good agreement with
numerical simulations as long as the temperatures were sufficiently 
low.\cite{wintel95}
Only at higher temperatures, a deviation from the  NL$\sigma$M predictions
for the temperature dependence of the correlation length was observed
which was attributed
to the appearance of isolated topological defects. In the numerical
simulations the high
temperature region showed some resemblance to the high
temperature region of XY-models \cite{wintel95} which 
indicates
that this region is characterized by free defects.
However, at present a good understanding of the
influence of such defects in non-collinear systems is still
lacking.\cite{kawamura98}

The topological defects of spirals have their origin in the chiral degeneracy
of the spiral, i.e. the spiral can turn clock- or anti-clock 
wise.\cite{kawamura98} 
At a topological defect, the spiral changes its chirality. As the 
chirality takes only two possible values, the defects are
 $Z_2$ defects.

It is then straightforward to find topological defect solutions of the
saddle point equations of a clean spiral.\cite{wintel94} 
The saddle point equations can be obtained from the perturbative expansion
of the energy density, Eq.~(\ref{pertexp}-\ref{pertexp4}).
One finds that extremal solutions must satisfy for each $j=x$,$y$,$z$ the
equations 
\begin{eqnarray}
\label{saddlepoint}
\left(1+b \delta_{jz}\right)\partial_\mu A_\mu^j
=2 b \epsilon_{zjk}A_\mu^z A_\mu^k,
\end{eqnarray}
where $j$ is not summed over. 
For $b>-1$ one finds solutions
of the form \cite{wintel94}
\begin{equation}
\label{topog}
g_s({\bf x})=\exp \left( \frac{i}{2} m^a \sigma^a \Psi({\bf x}) \right),
\end{equation}
where $\bf m$ is a space independent unit vector and $\Psi({\bf x})$ a
scalar function. With this Ansatz, one
has $A_\mu^i({\bf x})=\frac{1}{2} m^i \partial_\mu \Psi({\bf x})$ and thus,
upon insertion into Eq.~(\ref{saddlepoint}), one finds 
for ${\bf m}$ and $\Psi$ the equations ($j$ is again not summed over)
\begin{eqnarray}
\label{mpsi}
\left(1+b \delta_{jz}\right) m^j \partial_\mu^2 \Psi({\bf x})
=b \epsilon_{zjk}m^z m^k \left(\partial_\mu \Psi({\bf x})\right)^2.
\end{eqnarray}
The weight of the configuration described  by $g_s$ is given by (we set
$t_\mu=t_s$)
\begin{eqnarray}
\nonumber
H\left[ g_s \right] &=& \frac{1}{t_s} 
\int d^2{\bf x}\ \left[ {\bf A}_\mu^2
+ b \left(A_\mu^z\right)^2 \right] \\ 
&=&\frac{1}{4 t_s} \left[1+b \left(m^z \right)^2
\right]
\int d^2{\bf x}\ \left( \partial_\mu \Psi \right)^2.
\label{topener}
\end{eqnarray}
We see that for $b<0$, the energy is minimized for 
$\left(m^z\right)^2=1$ whereas
for $b>0$ the vector ${\bf m}$ is preferably orientated within the 
$x$-$y$ plane
with $m^z=0$. For both cases, Eqs.~(\ref{mpsi}) reduce to the two dimensional
Laplace equation 
${\bf \nabla}^2 \Psi({\bf x})=0$. This equation allows for topological
defect solutions with 
$\Psi(x,y)=\arctan(y/x)$. 
In the top of Figs.~\ref{topopic} and \ref{topopic2}
the spin distribution around isolated defects 
is shown for both $b<0$ and $b>0$.
Using Eq.~(\ref{topener}) one finds that
the energy of a topological defect solution 
$\Psi(x,y)$
diverges logarithmically
with the linear system size $R$, 
\begin{figure}[hbt]
\begin{center}
\includegraphics[width=6cm]{fig7a.epsi}
\end{center}
\begin{center}
\includegraphics[width=6cm]{fig7b.epsi}
\end{center}
\caption[]{\label{topopic} Single topological defect (top) and topological
defect pair (bottom) of a spiral with $b\le0$ (small scale AF 
fluctuations are not shown).}
\end{figure}
\begin{figure}[hbt]
\begin{center}
\includegraphics[width=6cm]{fig8a.epsi}
\end{center}
\begin{center}
\includegraphics[width=6cm]{fig8b.epsi}
\end{center}
\caption[]{\label{topopic2} Single topological defect (top) and topological
defect pair (bottom) of a spiral with $b\ge0$
(small scale AF 
fluctuations are not shown).}
\end{figure}
\begin{equation}
\label{topenergy}
\beta E=\frac{1+ (m^z)^2 b}{2t_s} \pi \ln R . 
\end{equation}

Because of this logarithmic divergence of the energy, isolated defects are
not present in absence of disorder and at sufficiently low temperatures.
It can also be shown,\cite{wintelphd}
that a bound state of defect pairs, described
by $g=g_{s1} g_{s2}$ with $g_{s1,2}=\exp \left[\frac{i}{2} {\bf m}_{1,2}
\cdot \fett{\sigma} \arctan\left(\frac{y-y_{1,2}}{x-x_{1,2}}\right) \right]$,
has a finite energy if ${\bf m}_1+{\bf m}_2=0$. 
Therefore, while isolated defects may be absent, defect pairs
will be present at any finite temperature. Figs.~\ref{topopic} and \ref{topopic2}
(bottom) show such a pair of topological defects for $b<0$ and $b>0$,
respectively.

This situation is reminiscent
of the one encountered in the XY model where at low temperatures also only
defect pairs are present. The 
pairs unbind at the critical Kosterlitz-Thouless temperature.
An unbinding of defects at a critical temperature or critical
disorder strength is also expected in the
present model. 
The topological
defects of the spiral differ however in important aspects from
those of the XY model. Spiral defects have a $Z_2$ charge while
XY defects have $Z$ charges. 
More importantly, as the present model possesses asymptotic
freedom, it has a finite correlation length $\xi$ at any finite temperature
even in absence of free defects. This implies that the 
logarithmic divergence in Eq.~(\ref{topenergy}) appears only up to a scale
$R<\xi$.
It is therefore not clear how a 
defect-unbinding would affect the system. A transition from a phase with
algebraically   decaying spin correlations  to a phase which shows an 
exponential decay,
as occurs in XY models, is clearly ruled out.
While in XY models topological defects can be relatively easily 
incorporated into
the analysis because they can be decoupled from the spin waves, this is
not the case for frustrated Heisenberg models.
If fluctuations
around the saddle point solution are taken into account, the defects
of spirals 
couple to the spin waves already at second order in an expansion in
the fluctuations \cite{wintel94}.
These difficulties have to date prevented a good understanding of
defect unbinding in frustrated systems.

A comparison to XY models is nonetheless quite illuminating. The kind
of disorder coupling we have used for the spiral phase is 
closely related in spirit to XY models with randomly fluctuating
phases, where the disorder is also introduced in the form of a
fluctuating gauge.\cite{scheidl97} 
If one ignores vortices, the influence of the disorder was shown to
amount to a simple renormalization of the spin stiffness, at all
orders in a perturbative treatment of the disorder coupling 
\cite{fisher85,scheidl97}
and no disordering transition as a function of the disorder strength
is found. However, once topological defects are included in the
analysis, the coupling of vortices to the random gauge field can lead
to a disordered phase even at $T=0$. This 
transition is driven by 
the creation of unpaired
defects if the fluctuations of the gauge
field are stronger than some critical value.\cite{scheidl97,nattermann95}
The critical disorder strength beyond which 
such defects appear can be estimated quite accurately when one
calculates the free energy of an isolated defect in presence
of disorder.\cite{scheidl97,cha95} It turns out that a similar
analysis of a single defect in a spiral in presence of disorder can
be carried out with some modifications, at least at the level
of saddle point solutions. Within this approximation,
the free energy of an isolated spiral defect is given by
\begin{equation}
\beta F=\frac{1+ (m^z)^2 b}{2t_s} \pi \ln R - \left[\mbox{ln} Z_d\right]_D ,
\end{equation}
where the second term contains the corrections due to the disorder coupling,
\begin{equation}
Z_d=\int d^2{\bf y} \exp \left( - 2 \int d^2{\bf x}\  p_k \ \epsilon_{ijk} \
\epsilon_{abc}\  A_\mu^i \  n_j^a \  n_k^c  \ Q_\mu^b \right)
\end{equation}
with ${\bf A}_\mu$, ${\bf n}_k$ obtained from 
Eqs.~(\ref{su2n}),~(\ref{su2a}) and (\ref{topog}).
With use of the replica trick 
$\left[\mbox{ln} Z_d\right]_D=
\lim_{N\to 0}\frac{1}{N} \ln \left[Z_d^N\right]_D$,
we have, assuming $b<0$,
\begin{eqnarray} \nonumber
\left[Z_d^N\right]_D= \int d^2{\bf y}_1 \dots d^2{\bf y}_N
\exp \left( 2 \lambda p_1^2\sum_{n,n'=1}^N \int  d^2{\bf x} \right. \\ \nonumber
\left. \partial_\mu \Psi_n \partial_\mu \Psi_{n'} \right);  
\end{eqnarray}
with $\Psi_n({\bf x})=\Psi({\bf x}-{\bf y}_n)$.
We write 
\begin{equation}
\int  d^2{\bf x}
\partial_\mu \Psi_n \partial_\mu \Psi_{n'}= -\frac{1}{2} \Delta_{nn'}
+ V^2  , 
\end{equation}
with $V^2\simeq 2 \pi \mbox{ln} R$ and $\Delta_{nn'} \simeq 4 \pi \mbox{ln}
|{\bf y}_n - {\bf y}_{n'}|$.\cite{nattermann95} For large separations
$|{\bf y}_n - {\bf y}_{n'}|$ we approximate $\Delta_{nn'}\simeq
4 \pi \mbox{ln} R$ while for small distances $\Delta_{nn'}$ is
negligible.
To find the highest
weight configuration, the replicas are grouped together in $N/m$ sets
containing each  $m$ replicas, with small distances between replicas
within a set and large distance for replicas in different sets. 
$\left[Z_d^N\right]_D$ then scales with $R$ as
\begin{equation}
\left[Z_d^N\right]_D \sim R^{4 \lambda \pi p_1^2 N^2 
+ \mbox{max}_m \left( 2 \frac{N}{m} 
-4 \pi \lambda p_1^2 N(N-m) \right)} . 
\end{equation}
In the limit $N \to 0$, maximization is replaced by minimization with respect
to $m$ in the range $0\le m \le 1$, so 
\begin{equation}
\beta F= \left[ 2 p_1 \pi - \mbox{min}_{0 \le m \le 1}
\left(2/m + 4 \lambda p_1^2 \pi m \right) \right] \ln R \ .
\end{equation}
For $2 \lambda p_1^2 \pi < 1$ one finds 
$\beta F= 2 [ p_1 \pi (1-2 \lambda p_1) -1]\ln R$ 
so that for $p_1 \pi (1 - 2 \lambda p_1)\leq 1$ free defects are
favorable. This is the phase boundary for thermal creation of defects.
At low temperatures, $2 \lambda p_1^2 \pi > 1$, one obtains
$\beta F= 2 \pi p_1 (1-\sqrt{8 \lambda/\pi}) \ln R$ and a
critical disorder strength $\lambda_c=\pi/8$ beyond which the disorder
favors isolated defects even at $T=0$. Similar considerations for the
case $b\ge 0$ lead to the same critical disorder strength and
the condition $\pi (p_1 + p_3) [1-\lambda (p_1 + p_3)]\leq 2$ for
thermal creation of free defects.

\begin{figure}[htb]
\begin{center}
\includegraphics[width=4cm]{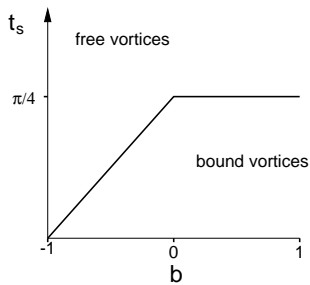}
\end{center}
\caption[]{\label{thermaldefects} The critical line for the thermal unbinding
of topological defects is shown in $b$, $t_s$ space.}
\end{figure}

Let us first discuss the results for the disorder free case $\lambda=0$.
The situation is summarized in Fig.~\ref{thermaldefects}, which shows
the line separating the regime where free vortices exist from the
regime in which all defects are bound. Notice that the unbinding
temperature goes linearly to zero in the limit $b\to-1$. At
$b=-1$, free defects are present at any finite
temperature. This is expected, as at $b=-1$ and finite $t_s$,  
the topological defects we discuss
here lose their meaning as the stiffness for rotations around the
collinear ordering axis disappears and the model becomes a 
$O(3)/O(2)$ model which has no finite temperature transition.
Whether or not free defects are present exactly at the point $b=-1$, $t_s=0$
depends on how this point is approached. To see this, we note that
the symmetry of the model in the limit $p_3\to \infty$ but finite
$p_1$ reduces to an XY symmetry as fluctuations of
the ${\bf n}_3$ vector get suppressed which forces all fluctuations
of the orthonormal pair ${\bf n}_{1,2}$ to lie within a plane. Therefore
one obtains an XY model with stiffness $p_1$. In terms of the $b$, $t_s$
parameters, this limit is approached as $t_s\to 0$ and $b\to -1$ with
finite $(1+b)/t_s=4 p_1$. Thus, depending on whether one
approaches the point $1+b=t_s=0$ with a slope
larger or smaller than the critical one given by
$(1+b)/t_s=4/\pi$, one arrives at the disordered phase or the ordered
phase of the XY model. This behavior is correctly reproduced by the 
free energy argument. The validity of the critical curve
$(1+b)/t_s=4/\pi$ also for finite $1+b>0$ 
is at least plausible, as topological defect solutions
also survive in this limit. Below this line, the
RG Eqs.~(\ref{RGT},~\ref{RGb}) hold and the system should scale
towards the point $b=0$. We can only speculate however what happens
above that line.
At least for some finite regime
near $b=-1$ the unbinding transition would presumably drive
$p_1$ to zero, as it does in the XY model, and affect the renormalization 
of $p_3$ only weakly. Thus, the appearance
of free defects will probably modify the
RG equations at high temperatures in such a way that the system will flow
back to the collinear point $b=-1$ as long as $1+b$ remains small enough.
For larger $b$ the nature of the RG is unclear. Numerical
simulations on triangular Heisenberg models \cite{southern93,southern95} 
have found however clear evidence for a defect unbinding transition.
As the triangular Heisenberg model is believed to have initially 
$b=1$,\cite{apel92} 
it is likely that an unbinding transition indeed occurs for every initial
value of $b$.
As no RG equations are available which can describe the transition,
the form of the correlation length near this transition is
unknown. It was however argued \cite{wintel94} that the temperature
dependence of the correlation length should cross over from the
NL$\sigma$M behavior to an XY behavior when the defects unbind.
Numerical results seem to support such a scenario.\cite{wintel95}

Let us now turn to the case with disorder.
Disorder will lead to the formation of free defects if $\lambda>\pi/8$.
According to the free energy argument above, this critical disorder
strength is independent of the stiffnesses $p_k$ and is thus 
also valid in the XY limit discussed above. 
For strong enough disorder, free 
topological defects will exist already at $T=0$, invalidating our
NL$\sigma$M analysis and producing very short low-temperature
correlation lengths for the spiral. 
For XY models,
the correlation length at $T=0$  behaves like 
$\xi \propto \exp (B/\sqrt{\lambda-\lambda_c})$ (with some constant $B$)
 near the critical 
disorder strength.\cite{scheidl97} 
This form of the correlation length has
a divergence of $\xi$ at $\lambda=\lambda_c$ which
 cannot be correct for the spiral because, as discussed above, 
even without vortices, the coupling of any finite amount of disorder
to the spins will lead to a finite correlation length. The correct
dependence of the correlation length at $T=0$ on the disorder 
is expected to be an interpolation between the NL$\sigma$M result
and the XY behavior. 

Certainly, the free energy argument is not expected to work as
well in the present model as it does for XY models. The parameters
$\lambda$ and $t_s$ flow to strong coupling and thus the 
predictions of the free energy argument also become scale
dependent. In other words, while at some small scale the
system might look stable against the creation of free defects,
at some larger scale the system will become unstable according
to the free energy argument. There does not seem to be a simple 
answer as to 
which scale is the correct one for applying the argument.
Note that such problems do not arise in the XY model where the
stiffness remains unchanged under the RG as long as vortices
are ignored. In view of the divergence of the $\lambda$ and $t_s$
parameters in the NL$\sigma$M, one possible scenario would be
that free defects will always be present at sufficiently large
length scales. Numerical results do however not support such
a scenario and rather point to the existence of a finite critical 
temperature.\cite{southern95} Below we shall apply the free
energy argument with the bare parameters, i.e. at the smallest
possible scale, which, if anything, would overestimate the stability of the
system against free defect formation.

\section{Comparison with experiments} 
\label{compexp}

Let us now compare our results with experimental data on
the SG phase of $\rm La_{2-x}Sr_xCuO_4$. Neutron scattering
data \cite{matsuda} have revealed an incommensurability of the spins which
scales roughly linearly with $x$. At very small $x$, a small deviation from 
the linear dependence is observed. Both features can be explained 
within the dipole model. The linear scaling is reproduced if the
fraction of the dipoles which are ordered is doping independent, i.e., 
the number of ordered dipoles scales linearly with doping.
The deviation from linearity might be explained with the increase
of the average separation between dipoles at small $x$ and a
resulting diminished tendency of the dipoles to align.

The same experimental data also shows the strong one dimensional
character of the IC modulation, i.e. the incommensurability
is observed only in one diagonal of the Cu-lattice ($b$-direction)
and thus breaks the symmetry of the square lattice. 
This phenomenon is usually interpreted as being due to the existence
of charge and spin stripes running along the other diagonal ($a$-direction).
However, this IC is also expected for a spiral along the $b$-direction
because its chirality breaks the translation symmetry (it can spiral
clock- or anticlock-wise). In addition, this symmetry breaking
is expected to show long-range order because the dipoles prefer
a discrete set of lattice orientations. 

Another important consequence of the spiral chirality is the
formation of topological defects. 
To judge, whether or not topological defects play a role in the
LSCO SG phase, we need an estimate of $\lambda$.  
We can use as a lower bound for $\lambda$ the result obtained
from the collinear analysis \cite{cherepanov99} where 
a disorder parameter equivalent to ours, but defined on
the much smaller scale of the AF unit cell, was used. 
>From a fit
of the $x$ dependence of the correlation length at $x<0.02$ and large
temperatures $T>T_N$, one obtains $\lambda\simeq 20 x$. 
In this regime
of $x$, the low temperature phase has long-range AF order and
a collinear analysis is well justified. 
We assume that the linear dependence of the disorder parameter on $x$,
$\lambda\simeq 20 x$,
also holds in the SG regime. This view is supported by measurements, 
which found that 
the width of the distribution of  internal magnetic
fields (i.e. local staggered moments) 
increases simply linearly with doping, with no detectable
change on crossing the AF/SG phase boundary,\cite{niedermayer98} see
also Fig.~\ref{niederm}.
It is remarkable that with
our above estimate for the critical disorder strength $\lambda_c=\pi/8$
we find a critical doping concentration $x_c\sim 0.02$. 
Considering
that $\lambda\simeq 20 x$ is a conservative lower bound of $\lambda$ 
at the long length scales relevant to spirals,
we conclude that in the entire SG phase, free topological 
defects will be present already at $T=0$, leading to a strongly 
disordered spiral phase. Experiments have in fact shown that
the correlation lengths in the SG regime are rather short
and of the same order as the periodicity of the IC
modulation.\cite{matsuda}
While this is in accordance with
the expected presence of topological defects, the correlation
lengths are so short that the condition $\xi \gg \left|{\bf q}_s\right|^{-1}$
is not fulfilled. The regime where spiral correlations become
dominant is therefore barely reached, and the RG scaling
predictions cannot be well tested. 

While qualitatively the experimental data supports a description of the
SG phase as a strongly disordered spiral state, both the extremely short
correlation lengths and our limited understanding of topological defects
prevent a more quantitative comparison.

However, our suggestion that the incommensurability 
of the spins is
related to ordered dipolar frustration centers can be directly 
tested experimentally on co-doped samples 
$\rm La_{2-x}Sr_{x}Zn_{z}Cu_{1-z}O_4$. 
Zn replaces  Cu in the CuO$_2$ planes and effectively removes one
spin. Zn doping leads  therefore to a dilution of the AF
but does not introduce frustration. Dilution is not very effective
in destroying the AF order and pure Zn doping (with $x=0$) 
leads to a destruction of long-range order only at
percolation
threshold that occurs for $x \approx$ 41\%. \cite{vajk02}
Surprisingly, for very small Sr concentration $x\leq 0.02$ it was found
that co-doping with Zn can increase $T_N$.\cite{huecker99} This
is remarkable as both kinds of impurities lead to a reduction of $T_N$
in singly doped samples. 
A possible explanation for
this behavior was suggested by Korenblit {\em et al.}\cite{korenblit99}
They put forward an argument, that  Zn impurities, if placed close
enough to the localized hole state, will destroy the frustrating nature
of the hole bound state. While their microscopic picture of frustration
is a classical one, a Zn impurity is also expected to strongly influence
the properties of the bound hole state within 
a more realistic quantum mechanical picture of frustration.
Although Zn couples only weakly to the spin degrees of freedom, 
if placed near a Sr donor, it disturbs the symmetry around 
the Sr atom and modifies the nature of the bound hole state.
As the Zn impurity breaks the sublattice pseudo-spin degeneracy of the
bound hole, the orientation of the dipole moment is no longer annealed
but becomes quenched. 
Another effect of the
breaking of the sublattice symmetry is that the weight of the bound hole
wave function near the wavevector $(\pi/2,\pi/2)$ or equivalent points
will be reduced. As it is these wavevectors which are responsible for
frustrating the spin background, one would expect a reduction or possibly
a complete destruction of the frustration caused by the hole.
Hence, the effective density of dipoles will be
renormalized to $x\rightarrow x (1-\gamma  z)$ where $\gamma$ must be
calculated from a microscopic theory (experiments indicate that
$\gamma$ is of order 2).\cite{korenblit99} 
Co-doping with
Zn then has two effects: First, it lowers the amount of frustration
in the sample and thus increases the correlation 
length, which would explain the experimentally observed increase
of $T_N$ with z for $x=0.017$.\cite{korenblit99,huecker99} 
Furthermore, the effect of quenching the dipole moments will be
the same as destroying them altogether with respect to the
incommensurability, as the incommensurability is determined solely
by the ordered moments.
Thus, co-doping with Zn will
lead to a decrease of the incommensurability by a factor
$1-\gamma z$. In contrast, within a stripe picture, co-doping
with Zn is not expected to change the incommensurability as
the hole density is not affected by Zn doping. 
Previous
measurements in the superconducting phase ($x = 0.12$ and $x = 0.14$),
where the stripe model is believed to be valid, have shown that the 
incommensurability indeed remains intact upon co-doping with 
Zn.\cite{Hiro,Yama2,Kimu}. 
Within a stripe picture, the only effect of Zn co-doping in the SG
regime should be pinning of stripes, which would lead to a reduced
correlation length.\cite{us} 
Therefore, neutron scattering experiments within the SG regime
of Zn co-doped samples
could clarify the debate, if the magnetic incommensurability observed in
the SG regime is to be interpreted within a stripe or a frustration
based model. 

It is interesting that symmetry arguments similar to those
just used to discuss Zn co-doping also give a simple
explanation for the absence of any incommensurate signal in Li doped 
La$_{2}$Cu$_{1-y}$Li$_{y}$O$_4$. For small $y$, these compounds show
a magnetic phase diagram which is almost identical to Sr doped 
samples\cite{sasagawa02}
with the notable exception that the magnetic correlations always
remain commensurate.\cite{bao03} 
Like Sr, each Li atom introduces an excess hole
in the CuO$_2$ plane which, at least for small doping concentrations,
remains weakly localized to its dopant. The important difference
is that Li replaces Cu in the crystal and thus has a different symmetry
with respect to the magnetic sublattice ordering than a Sr hole. Specifically,
the sublattice position of the Li atom breaks the pseudospin
degeneracy present in Sr doped samples. Assuming that otherwise the
origin of frustration is the same, the only difference between Sr and Li
doped samples is that the dipole moment assigned to the Li bound hole
is quenched, whereas the one of the Sr hole is annealed. Thus, ordering
of these moments and the development of incommensurate 
correlations cannot occur
in Li doped samples. 

In conclusion, we have presented a detail picture of the dipole model
of frustration and discussed its applicability to the weakly doped
regime of cuprate materials. 
Most of the key characteristics
of these materials were already known to be in accordance with the
model and we showed that
incommensurate correlations appear also naturally within the dipole
picture. We extended the commensurate model to allow for a description
of the resulting disordered spiral spin phases. Finally, we suggested 
an experiment which would allow to verify
whether the frustration based dipole model
or the stripe picture is realized within the weakly doped regime 
of cuprates.

\section{Acknowledgments}

We acknowledge fruitful discussions with V.\ Gritsev, V.\ Juricic,
B.\ Normand, and B.\ Simovic. 
C.\ M.\ S.\ is supported by the Swiss National Foundation under
grant 620-62868.00.

\begin{appendix}
\section{SU(2) representation}
\label{appsu2}
The orthonormal basis ${\bf n}_k$ can be related to an element
$g$ of $SU(2)$ through 
$g \sigma^k g^{-1}={\bf n}_k \cdot$ \boldmath$\sigma$\unboldmath, or
\begin{eqnarray}
n_k^a=\frac{1}{2} \mbox{tr} \left\{\sigma^a g \sigma^k g^{-1}
\right\}
\end{eqnarray}
For the derivative one finds, using $\partial_\mu \left(g g^{-1}\right)=0$,
\begin{eqnarray} \nonumber
\partial_\mu n_k^a &=& \frac{1}{2}\mbox{tr} \left\{\sigma^a 
\partial_\mu g \sigma^k g^{-1}
+ \sigma^a g \sigma^k \partial_\mu g^{-1} \right\}  \\
&=& \frac{1}{2}\mbox{tr} \left\{ \sigma^k \left[ g^{-1} \sigma^a g,\ g^{-1}
\partial_\mu g \right] \right\}.
\end{eqnarray}
Introducing $g^{-1}\partial_\mu g = i {\bf A}_\mu \cdot$ 
\boldmath$\sigma$ \unboldmath and with 
$\left[ \sigma^i, \ \sigma^j\right] = 2 i \epsilon_{ijk} \sigma^k$
one finds
\begin{equation}
\partial_\mu n_k^a = 2 \epsilon_{ijk} A_\mu^i n_j^a.
\end{equation}
Therefore, we have (with $p_{1\mu}=p_{2\mu}$)
\begin{eqnarray}
p_{k\mu} \left( \partial_\mu {\bf n}_k \right)^2
&=& 4 p_{k\mu} \left( \epsilon_{ijk} A_\mu^i n_j^a \right)^2
=4 p_{k\mu} \left( \epsilon_{ijk} \right)^2 \left(A_\mu^k\right)^2 \nonumber \\
&=& \frac{2}{t_{\mu}} \left[ {\bf A}_\mu^2 + b \left( A_\mu^z \right)^2 \right],
\end{eqnarray}
with $t_{\mu}^{-1}=2 (p_{1\mu} + p_{3\mu})$ and 
$bt_{\mu}^{-1}= 2 (p_{1\mu}-p_{3\mu})$.

\section{Expanding the energy functional in $\varphi^i$}
\label{expphi}

To do the RG, we introduce 
$g= \tilde{g} \exp (i$\boldmath$\varphi \cdot \sigma$
\unboldmath$)$, where $\varphi^a$ are fast fields fluctuating 
with wavelengths $[\Lambda^{-1} ,1]$ and $\tilde{g}$ 
has only slow fluctuations in
the range $[0,\Lambda^{-1} ]$.
For the 1-loop calculation, we need
to expand ${\bf n}_k$ and ${A}_\mu^k$ up 
to second order in $\varphi^a$.
We then find
\begin{eqnarray}
n_i^a&=& 
\frac{1}{2} \mbox{tr}\left\{ \sigma^a \tilde{g} \exp \left(i 
\mbox{\boldmath $\varphi \cdot \sigma$ \unboldmath} \right) \sigma^i
\exp \left(-i 
\mbox{\boldmath $\varphi \cdot \sigma$ \unboldmath} \right) \tilde{g}^{-1}
\right\} \nonumber \\
&=&
\tilde{n}_i^a + \frac{i}{2} \mbox{tr} \left\{ \sigma^a \tilde{g} \left[
\mbox{\boldmath $\varphi \cdot \sigma$\unboldmath}, \sigma^i \right] 
\tilde{g}^{-1}
\right\} \nonumber \\
&+& \frac{1}{2} \mbox{tr} \left\{ 
\sigma^a \tilde{g} \left( \mbox{\boldmath $\varphi \cdot \sigma$ \unboldmath}
\sigma^i \mbox{\boldmath $\varphi \cdot \sigma$ \unboldmath} 
- \frac{1}{2} 
\left(\mbox{\boldmath $\varphi \cdot \sigma$ \unboldmath}\right)^2
\sigma^i \right. \right.
\nonumber  \\   
&-& \left. \left. \frac{1}{2} \sigma^i \left(
\mbox{\boldmath $\varphi \cdot \sigma$ \unboldmath}\right)^2
\right) \tilde{g}^{-1} \right\}  + {\cal O}(\varphi^3)
\nonumber \\ &=&
\tilde{n}_i^a +2 \epsilon_{ijk} \varphi^j \tilde{n}_k^a 
+ \varphi^j \varphi^k R_{jk}^{ai}
+ {\cal O}(\varphi^3),
\end{eqnarray}
where
$$
R_{jk}^{ai}= \frac{1}{2} \ \mbox{tr} \left\{ \sigma^a \tilde{g}
\left( \sigma^j \sigma^i \sigma^k -\frac{1}{2} \sigma^j \sigma^k
\sigma^i - \frac{1}{2} \sigma^i \sigma^j \sigma^k \right)
\tilde{g}^{-1} \right\}.
$$
It turns out, that in the RG we will only need the diagonal
components of $R_{jk}^{ai}$ with $j=k$ which have the much simpler
form 
$R_{zz}^{ai}= -2 \left(\epsilon_{zqi}\right)^2 \tilde{n}_{i}^{a}$
(we put here $j=k=z$ to make clear that $z$ is not a silent
index, the equation also holds for $j=x,y$).
Similarly, we find
\begin{eqnarray}
A_\mu^i&=& 
\frac{1}{2i}\ \mbox{tr} \left\{
\sigma^i \exp \left(-i \mbox{\boldmath $\varphi \cdot \sigma$\unboldmath}
\right) \left[ \partial_\mu +\tilde{g}^{-1}\partial_\mu \tilde{g} \right]
\exp \left(i \mbox{\boldmath $\varphi \cdot \sigma$\unboldmath} \right)
\right\} \nonumber \\
&=& \tilde{A}_\mu^i +  \frac{1}{2} \mbox{tr}
\left\{\sigma^i \left( \partial_\mu \fett{\varphi \cdot \sigma}
+ \frac{1}{2i} \left[ \fett{\varphi \cdot \sigma}, 
\partial_\mu\fett{\varphi\cdot \sigma} \right]  \right. \right.\nonumber \\
&+& \left. \left.
i \left[\tilde{\bf A}_\mu\fett{\cdot \sigma}, 
\fett{\varphi \cdot \sigma} \right]
+ \fett{\varphi \cdot \sigma}\ \tilde{\bf A}_\mu\fett{\cdot \sigma}\
\fett{\varphi \cdot \sigma}
\right. \right. \nonumber \\  
&-& \left. \left.
\frac{1}{2} \left(\fett{  \varphi \cdot \sigma}\right)^2 
\tilde{\bf A}_\mu\fett{\cdot \sigma}- \frac{1}{2} 
\tilde{\bf A}_\mu\fett{\cdot \sigma}
 \left(\fett{  \varphi \cdot \sigma}\right)^2
\right) \right\} + {\cal O}(\varphi^3) \nonumber \\
&=& \tilde{A}_\mu^i + \partial_\mu \varphi^i 
+\epsilon_{ijk} \varphi^j \partial_\mu \varphi^k
+2 \epsilon_{ijk}\varphi^j \tilde{A}_\mu^k
 -2 \tilde{A}_\mu^i\ \fett{\varphi}^2
\nonumber \\ 
&+& 2 \tilde{\bf A}_\mu\fett{\cdot \varphi} \ \varphi^i+ {\cal O}(\varphi^3).
\end{eqnarray}

\section{Propagator of the $\varphi^i$ fields}
\label{propphi}

As already mentioned, there is a small spatial anisotropy
in the stiffnesses $p_{k \mu}$, i.e. $p_{k1}\neq p_{k2}$.
We shall keep here the spatial dependence of the stiffnesses
$p_{k\mu}$ up to first order in the anisotropy, assuming
that the anisotropy $\kappa$, which we define through
$p_{k1}/p_{k2}=1+\kappa$, is independent of the $k$ index. Thus we can
absorb the anisotropy into the $t_{\mu}$ parameter while 
$b$ remains isotropic. We then define $t_s=\sqrt{t_1 t_2}$ and
$t_{1,2}\simeq (1 \pm \kappa/2) t_s$. For future use, we also
define the isotropic stiffnesses $p_k=\sqrt{p_{k1}p_{k2}}$. 
It is not clear whether the isotropy of $b$ is preserved under
the RG and we have made no attempt to write down the RG equations
in presence of anisotropy. In principle, if 
$b$ remains isotropic, the results obtained
below allow to determine the flow of the anisotropy parameter
$\kappa$ under the RG.
For possible future use, we will therefore keep
the perturbative expansion with the anisotropy. The
results used in the body of this work have however been obtained
for an isotropic $t_\mu$=$t_s$, i.e. $\kappa=0$.

We need to expand the exponential $\exp(-{H_P})$ and
integrate out the $\varphi^i$ fields. Taking the average over
the $\varphi^i$ fields is done with the Gaussian term $H_\varphi$ of
Eq.~(\ref{pertexp}).
The propagator for the $\varphi^i$ is thus quite simple and
becomes, to lowest order in the anisotropy $\kappa$
\begin{eqnarray}
C^i({\bf x})&&:= \left< \varphi^i({\bf x}) \varphi^i({\bf 0})\right>_\varphi
=\frac{t_s}{2(1+b \delta_{iz})}
\int \frac{d^2 {\bf k}}{(2 \pi)^2} 
\frac{e^{i {\bf k}\cdot {\bf x}}}{k^2} \nonumber \\ &&
\left( 1+ \kappa
\frac{k_1^2 - k_2^2}{2 k^2} \right) 
\times (\Upsilon(k,\Lambda)-\Upsilon(k,1)).
\end{eqnarray}
The IR cutoff is provided by the function $\Upsilon(k,\Lambda)$. A sharp
cutoff, $\Upsilon(k,\Lambda)=\Theta(k-\Lambda^{-1})$ has the disadvantage
of producing a long-ranged $C^i$ and we therefore adopt
instead $\Upsilon(k,\Lambda)=[1+ (k \Lambda)^{-2}]^{-1}$, which renders
$C^i$ short ranged.

In our RG calculation we will mainly need $C^i({\bf 0})$
which has the  form
$$
C^x({\bf 0})=C^y({\bf 0})=\frac{t_s}{4 \pi} \ln \Lambda 
 + {\cal O}(\kappa^2), \ \ \ \
C^z({\bf 0})=\frac{1}{1+b}C^x({\bf 0}).
$$
Another useful formula is
\begin{eqnarray}
\label{useful}
t_\mu^{-1} \int d^2{\bf x} \ \left(\partial_\mu C^x \right)^2
= \frac{1}{2} C^x({\bf 0}) + {\cal O}(\kappa^2).
\end{eqnarray}

\section{Renormalization}

We can immediately discard all
terms of third or higher power in $\tilde{\bf A}_\mu$ as these terms
are irrelevant in a RG sense. Terms second order in $\tilde{\bf A}_\mu$
 renormalize $t_\mu$ and
$b$, whereas terms linear in $\tilde{\bf A}_\mu$ are responsible for the
renormalization of the disorder variance $\lambda$.

First, we note that the terms $H_2$, $H_3$ do not contribute to
the renormalization, as was pointed out for the 
calculation of the RG for the disorder free system in Ref.\ [\cite{wintelphd}].
This is because these terms are linear in $\varphi$ while they do not
involve a disorder field ${\bf Q}_\mu$. 
 For an abelian
theory, such terms cannot
contribute because the fast $\varphi^i$ fields and the slow
$\tilde{\bf A}_\mu$ fields have their support in orthogonal parts of the
wave vector space. 
Here, for the non-abelian case, this argument is not sufficient because
the $\tilde{\bf A}_\mu$ fields are not linearly related to the fields
$g$. For the present non-abelian theory this is nonetheless true,
although an explicit calculation is required to see this.
For example, $H_{2}^2$ does not contribute, because its contribution
is built from terms of the form (we omit the upper $i$ indices 
of $C^i$ and $A_\mu^i$ here
for simplicity)
\begin{eqnarray} 
\label{zeroterm}
\int d^2{\bf x} \int d^2{\bf x}^\prime \ \tilde{A}_\mu({\bf x})
 \tilde{A}_{\mu^\prime}({\bf x}^\prime)
\partial_\mu \partial_{\mu^\prime} C({\bf x}-{\bf x}^\prime)
\end{eqnarray}
To evaluate this term, we change to center of mass ($\bf y$) and 
relative (${\bf y}^\prime$) coordinates and then perform a
gradient expansion in the relative coordinate.
Only the lowest order contribution is of interest, as higher
order terms involve a local coupling of the type 
$A_\mu \left(\partial_\nu \right)^n A_{\mu^\prime}$ 
with $n>0$ which are irrelevant from
a scaling point of view. The lowest order term is then
\begin{eqnarray}
-\int d^2{\bf y} \ \tilde{A}_\mu({\bf y})\tilde{A}_{\mu^\prime}({\bf y})
\int d^2{\bf y}^\prime 
\partial_\mu \partial_{\mu^\prime} C({\bf y}^\prime)
\end{eqnarray}
which vanishes because the last integral is zero. 
In the following we will 
omit $H_2$ and $H_3$ from the analysis, because terms involving
them do not contribute. This can be shown 
for each term in a way similar
to the one just shown.

We want to find the RG equations up to second order in 
$t_\mu$ and $\lambda$. In the $n$th order of the cumulant expansion of 
$\cal F$, Eq.~(\ref{cumulant}), 
we 
only need to consider terms 
which have a total number of $\fett{\varphi}$ and ${\bf Q}_\mu$
fields less than $2n+2$. This is  because each term of order 
$n$ carries a factors 
$t_s^{-n}$ from the prefactors of the terms in $H_p$
and each pair of $\fett{\varphi}$ (${\bf Q}_\mu$) produces
a factor $t_s$ ($\lambda$).

We begin first with the terms renormalizing $t_\mu$ and $b$,
where we give a detailed calculation only for the terms
up to second order in $H_p$. The calculation of higher order terms 
is quite lengthy although conceptually easy and we therefore just
present the results of the calculation.

\begin{widetext}
\subsection{Terms which renormalize $t_\mu$ and $b$}
\label{apptterms}

\subsubsection{First order in $H_p$}

There is only one term quadratic in 
$\tilde{\bf A}_\mu$ which contributes, $H_{4}$
(the $\varphi^i$ average over $H_3$ is zero).
\begin{eqnarray}
-\left<H_{4}\right>_{\varphi c} &=& 
-4 \frac{b}{t_{\mu}} \int d^2 {\bf x} \left[ \epsilon_{zjk} 
\epsilon_{zj^\prime k^\prime} \tilde{A}_\mu^k\tilde{A}_\mu^{k^\prime}
\left< \varphi^j \varphi^{j^\prime} \right>_\varphi - 
\left(\tilde{A}_\mu^z\right)^2 
\left< \varphi^l \varphi^{l} \right>_\varphi
+ \tilde{A}_\mu^z \tilde{A}_\mu^l 
\left< \varphi^z \varphi^{l} \right>_\varphi \right]\nonumber \\
&=&
-4 \frac{b}{t_{\mu}} \int d^2 {\bf x} \left[ \left(\epsilon_{zjk}\right)^2  
\left(\tilde{A}_\mu^k\right)^2 C^j({\bf 0}) - 
\left(\tilde{A}_\mu^z\right)^2 \sum_l C^l({\bf 0}) 
+ \left(\tilde{A}_\mu^z \right)^2 C^z({\bf 0})\right]\nonumber \\
&=&- 4 b \ t_\mu^{-1} \int d^2 {\bf x}
\left[\tilde{\bf A}_\mu^2 -
3 \left(\tilde{A}_\mu^z\right)^2\right] C^x({\bf 0}).
\end{eqnarray}

\subsubsection{Second order in $H_p$}

Terms with odd numbers of $\varphi^i$ or ${\bf Q}_\mu$ are zero after
performing the $\varphi^i$ and disorder average. There
are then only two terms we need to consider, $H_1^2$ and
$H_{c1}^2$ ($H_{c3}^2$ has a total of six $\varphi^i$ and $Q_\mu^i$
fields and does not contribute and $H_{2}$ terms do not contribute
as mentioned above).
For $H_1^2$ we have
\begin{eqnarray}
\frac{1}{2}\left[\left< H_{1}^2 \right>_{\varphi c} \right]_D &=&
\frac{1}{2}\left< H_{1}^2 \right>_{\varphi c} 
\\ &=&
2 t_\mu^{-1}t_{\mu^\prime}^{-1} 
\int d^2{\bf x} 
\ d^2{\bf x}^\prime
\ \tilde{A}_\mu^i({\bf x}) \tilde{A}_{\mu^\prime}^{i^\prime}({\bf x}^\prime) 
\epsilon_{ijk} \epsilon_{i^\prime j^\prime k^\prime} 
\left( 1- b \delta_{iz} + 2 b \delta_{jz}
\right) \nonumber \\ && \times
\left( 1- b \delta_{i^\prime z} + 2 b \delta_{j^\prime z}
\right)
\left<\partial_\mu \varphi^j({\bf x})
\varphi^k({\bf x}) \partial_{\mu^\prime}\varphi^{j^\prime}({\bf x}^\prime)
\varphi^{k^\prime}({\bf x}^\prime) \right>_\varphi 
\nonumber 
\end{eqnarray}
The four point average can be decomposed according to Wick's Theorem.
Nonzero contributions arise from the contractions $\left<jk^\prime\right>
\left<j^\prime k\right>$
and $\left< jj^\prime \right> \left< kk^\prime\right>$.
We again employ
an expansion of $H_1^2$ in the relative coordinate and keep
only the zeroth order term of the expansion. This yields
\begin{eqnarray}
\frac{1}{2}\left< H_{1}^2 \right>_{\varphi c} 
&\simeq& 2 t_{\mu}^{-2} \int d^2{\bf x}\tilde{A}_\mu^i \tilde{A}_\mu^{i^\prime}
\epsilon_{ijk} \epsilon_{i^\prime j^\prime k^\prime} 
\left( 1- b \delta_{iz}+ 2 b \delta_{jz} \right)
\left( 1- b \delta_{i^\prime z} + 2 b \delta_{j^\prime z}
\right) \nonumber \\ &&
\times \left(\delta_{jj\prime} \delta_{kk\prime}
- \delta_{kj\prime} \delta_{jk\prime} \right)
\int d^2{\bf y} \  \partial_\mu  C^j({\bf y})  \partial_\mu C^k({\bf y}) 
\nonumber \\
&=& 4 t_\mu^{-2} \int d^2{\bf x} \left(\tilde{A}_\mu^i\right)^2
\left(\epsilon_{ijk}\right)^2 \left(1-b \delta_{iz} +2b \delta_{jz}
\right) \left(1-b \delta_{iz} + b \delta_{jz} + b \delta_{kz}\right)
\nonumber \\ && \times
\int d^2{\bf y} \  \partial_\mu  C^j({\bf y})  \partial_\mu C^k({\bf y}).
\end{eqnarray}
With use of Eq.~(\ref{useful}), we finally find
\begin{eqnarray}
\frac{1}{2}\left< H_{1}^2 \right>_{\varphi c} =
2 t_\mu^{-1} \int d^2 {\bf x} 
\left[\tilde{\bf A}_\mu^2 (1+b) + \left(\tilde{A}_\mu^z\right)^2
b(b-3) \right] C^x({\bf 0})
\end{eqnarray}
The other second order contribution is
\begin{eqnarray}
\frac{1}{2}\left[ \left< H_{c1}^2 \right>_{\varphi c}\right]_D &=&
8 \int d^2 {\bf x} d^2 {\bf x}^\prime p_{k\mu}p_{k^\prime \mu^\prime}
\epsilon_{ijk}\epsilon_{i^\prime j^\prime k^\prime}
\epsilon_{abc}\epsilon_{a^\prime b^\prime c^\prime}
\left\{\epsilon_{klm}\tilde{n}_j^a \tilde{n}_m^c \tilde{A}_\mu^i
\right. \nonumber \\ && \left.
+\epsilon_{jlm}\tilde{n}_k^c \tilde{n}_m^a \tilde{A}_\mu^i
+\epsilon_{ilm}\tilde{n}_j^a \tilde{n}_k^c \tilde{A}_\mu^m \right\}
\left\{\epsilon_{k^\prime l^\prime m^\prime}
\tilde{n}_{j^\prime}^{a^\prime} \tilde{n}_{m^\prime}^{c^\prime}
 \tilde{A}_{\mu^\prime}^{i^\prime}
+\epsilon_{j^\prime l^\prime m^\prime}\tilde{n}_{k^\prime}^{c^\prime}
 \tilde{n}_{m^\prime}^{a^\prime} \tilde{A}_{\mu^\prime}^{i^\prime}
\right. \nonumber \\ && \left. 
+\epsilon_{i^\prime l^\prime m^\prime}\tilde{n}_{j^\prime}^{a^\prime}
 \tilde{n}_{k^\prime}^{c^\prime} \tilde{A}_{\mu^\prime}^{m^\prime}
 \right\} 
\delta_{l l^\prime} \ C^l({\bf x}-{\bf x}^\prime) 
\left[ Q_\mu^b({\bf x}) Q_{\mu^\prime}^{b^\prime}({\bf x}^\prime)
\right]_D.
\end{eqnarray}
Using $\left[ Q_\mu^b({\bf x}) Q_{\mu^\prime}^{b^\prime}({\bf x}^\prime)
\right]_D=\delta_{b b^\prime} \ \delta_{\mu \mu^\prime} \
\delta({\bf x}-{\bf x}^\prime) \lambda$, 
$\epsilon_{abc} \epsilon_{a^\prime b c^\prime}=
\delta_{a a^\prime} \ \delta_{c c^\prime}-\delta_{a c^\prime} 
\delta_{c a^\prime}$ and the orthonormality of the ${\bf n}_k$,
we find after some algebra
\begin{eqnarray}
\frac{1}{2}\left[\left< H_{c1}^2 \right>_{\varphi c} \right]_D&=&
2  \lambda\ b^2\ t_\mu^{-2}
\int d^2 {\bf x}
\left[\tilde{\bf A}_\mu^2 + \left(\tilde{A}_\mu^z\right)^2
\right] C^x({\bf 0}).
\end{eqnarray}
Higher order terms can be evaluated in much the same way
as the first and second order terms, although the large
number of indices makes their evaluation more tedious. We therefore
refrain
here from a detailed presentation of these terms and just state
the results.

\subsubsection{Third order in $H_p$}

Terms of second order in $\tilde{\bf A}_\mu^2$ are produced by
$\left(H_1+H_{c1}+H_{c3}\right)^2 (H_{c2}+H_{c4})$. However,
only the terms $H_1 (H_{c1} + H_{c3})(H_{c2}+H_{c4})$ have
even powers of ${\bf Q}_\mu$. Terms with eight or more  
${\bf \varphi}$ and ${\bf Q}_\mu$ fields again do not contribute
to second  order in $\lambda,t_\mu$.
Thus we are left with only
$H_{1} H_{c1}{H_{c2}}$ . We find
\begin{eqnarray}
-\left[ \left< H_1 H_{c1} H_{c2} \right>_{\varphi c} \right]_D
&=& - 2  \lambda t_\mu^{-2}\ b \int d^2 {\bf x}\left[\tilde{\bf A}_\mu^2  (1+b)
+ \left(\tilde{A}_\mu^z\right)^2 
\left( b-3 \right)
\right] C^x({\bf 0}).
\end{eqnarray}
We further need to consider terms of the type $(H_{c2} + H_{c4})^2 H_4$.
Only
$H_{c2}^2 H_4$ has less than eight
${\bf \varphi}$,${\bf Q}_\mu$ fields and even powers of both
fields. We find
\begin{eqnarray}
-\frac{1}{2}\left[ \left< H_{c2}^2 H_4 \right>_{\varphi c}  \right]_D
&=& -2 \lambda \ b \  t_s^{-1} \ t_\mu^{-1} 
\int d^2 {\bf x}\left[\tilde{\bf A}_\mu^2 - 3 
\left(\tilde{A}_\mu^z\right)^2 \right] C^x({\bf 0}).
\end{eqnarray}

\subsubsection{Fourth order in $H_p$}

Possible contributions arise from the terms 
$(H_1+H_{c1}+H_{c3})^2 (H_{c2}+H_{c4})^2$.
Discarding terms with ten or more $\varphi^i$,${\bf Q}_\mu$ fields,
we are left with $H_{c2}^2 H_1^2$ and $H_{c2}^2 H_{c1}^2$. However,
the connected part of the $\varphi^i$ average of $H_{c2}^2 H_{c1}^2$ is
zero (its finite disconnected parts enter the renormalization of
the disorder, see below), and the only contribution is therefore
\begin{eqnarray}
\frac{1}{4}\left[ \left< H_{c2}^2 H_1^2 \right>_{\varphi c}\right]_D
&=&  \lambda t_s^{-1} t_\mu^{-1} \int d^2 {\bf x}\left[\tilde{\bf A}_\mu^2
 (2+b) (1+b) 
+\left(\tilde{A}_\mu^z\right)^2  b\ (b-7)
\right] C^x({\bf 0}).
\end{eqnarray}
Terms of the form $H_4 (H_{c2}+H_{c4})^3$ do not contribute because
their disorder average is zero.
Higher order terms in $H_p$ do not contribute because they either
involve more than four ${\bf Q}_\mu$ terms and are therefore of
higher order than $\lambda^2$ or they do not contain finite connected
parts. For example, the term $\left<H_{4} H_{c2}^4\right>_{\varphi c}$ 
decomposes into products
of averages of $\left< H_4 \right>_{\varphi c}$ or 
$\left< H_4 H_{c2}^2\right>_{\varphi c}$ and  
$\left< H_{c2}^2 \right>_{\varphi c}$.

\end{widetext}

\subsection{Terms which renormalize $\lambda$}
\label{applterms}
To find the renormalization of the variance of the disorder  
distribution, we first collect all connected terms linear in 
$\tilde{A}_\mu^i$. 
We list the contributions order by order below. 

\subsubsection{First order in $H_p$}

Only three terms are linear in $\tilde{A}_\mu^i$, $H_1$, $H_{c1}$ and $H_{c3}$.
However, both $H_1$ and $H_{c1}$ have a zero $\varphi^i$ average
and only $\left< H_{c3} \right>_{\varphi c}$ contributes.

\subsubsection{Second order in $H_p$}

At second order there are contributions from $\left< H_{c1} H_{c2}
\right>_{\varphi c}$ and $\left< H_{1} H_{c4}
\right>_{\varphi c}$. There is no contribution to second order
in $\lambda$, $t_\mu$ of the disorder renormalization from
$\left< H_{c3} H_{c4}\right>_{\varphi c}$ because this term
has six $Q_\mu^i$, $\varphi^i$.

\subsubsection{Third order in $H_p$}

There are contributions from $\left< H_{c1} H_{c2} H_{c4}\right>_{\varphi c}$,
$\left< H_{c3} H_{c2}^2 \right>_{\varphi c}$ and 
$\left< H_{1} H_{c2}^2 \right>_{\varphi c}$. The terms
$\left< H_{c3} H_{c4}^2 \right>_{\varphi c}$ and 
$\left< H_{1} H_{c4}^2 \right>_{\varphi c}$ do not contribute, as they
contain  eight or more $Q_\mu^i$, $\varphi^i$ fields.

\subsubsection{Fourth order in $H_p$}

Only one term contributes,
$\left< H_1 H_{c2}^2 H_{c4}\right>_{\varphi c}$. All other terms
have ten or more $Q_\mu^i$, $\varphi^i$ fields or more
than three ${\bf Q}_\mu$ fields and thus do not contribute.
The same argument applies to all terms generated by higher order of $H_p$.

\begin{widetext}
\subsection{Calculating the renormalized disorder variance}
\label{applr2}
We now must calculate the variance of all terms at the new
length scale $\Lambda^{-1}$ which are linear in $\tilde{A}_\mu^i$. 
These are the terms just found above
plus $H_{c0}$. Thus, we need
to calculate the variance of
\begin{eqnarray} &&
-H_{c0} - \left<H_{c3}\right>_{\varphi c}
+ \left<H_{c1} H_{c2}\right>_{\varphi c} 
+ \left<H_{1} H_{c4}\right>_{\varphi c} 
- \left<H_{c1} H_{c2} H_{c4}\right>_{\varphi c} 
- \frac{1}{2} \left<H_{c2}^2 H_{c3}\right>_{\varphi c}\nonumber \\ &&
- \frac{1}{2} \left<H_{c2}^2 H_{1}\right>_{\varphi c} 
+ \frac{1}{2} \left<H_1 H_{c2}^2 H_{c4}\right>_{\varphi c}
\end{eqnarray}
To order $\lambda^2$, the following terms contribute to the
variance.
\begin{eqnarray}
\left[H_{c0}^2\right]_D&=& 
\lambda t_\mu^{-2} 
\int d^2{\bf x} \left\{ \left[ \left(\tilde{A}_\mu^x\right)^2
+ \left(\tilde{A}_\mu^y\right)^2 \right] +  \left(\tilde{A}_\mu^z\right)^2 (1+b)^2
\right\}, \nonumber \\ 
2 \left[ \left<H_{c3}\right>_{\varphi c} H_{c0} \right]_D
&=& 8 \lambda t_\mu^{-2} \int d^2{\bf x} \left\{ \left[ \left(\tilde{A}_\mu^x\right)^2
+ \left(\tilde{A}_\mu^y\right)^2 \right] b \right. \nonumber \\
&& \ \ \ \ \ \ \ \ \ \ \ \ \ \ \ \ \ \ \ \ \
- \left.
\left(\tilde{A}_\mu^z\right)^2 
2 b (1+b) \right\} C^x({\bf 0}), \nonumber \\ 
-2 \left[ \left<H_{1}H_{c4}\right>_{\varphi c} H_{c0} \right]_D &=&
-4 \lambda t_\mu^{-2} \int d^2{\bf x} \left\{ \left[ \left(\tilde{A}_\mu^x\right)^2
+ \left(\tilde{A}_\mu^y\right)^2 \right] (1+b)   \right. \nonumber \\ 
&& \ \ \ \ \ \ \ \ \ \ \ \ \ \ \ \ \ \ \ \ \
+ \left. \left(\tilde{A}_\mu^z\right)^2 
(1-b)^2 (1+b) \right\} C^x({\bf 0}), \nonumber \\ 
2 \left[ \left<H_{c1}H_{c2}H_{c4}\right>_{\varphi c} H_{c0} \right]_D
&=&  2 \lambda^2 t_\mu^{-3}  \int d^2{\bf x} 
\left\{ \left[ \left(\tilde{A}_\mu^x\right)^2 + \left(\tilde{A}_\mu^y\right)^2 \right]b (1+b)
\right. \nonumber \\ && \ \ \ \ \ \ \ \ \ \ \ \ \ \ \ \ \ \ \ \ \
+ \left. \left(\tilde{A}_\mu^z\right)^2 2 b (b^2 -1) \right\}  C^x({\bf 0}), 
\nonumber \\
\left[ \left<H_{c2}^2H_{c3}\right>_{\varphi c} H_{c0} \right]_D
&=& 
4 \lambda^2 t_\mu^{-2} t_s^{-1}  \int d^2{\bf x} 
\left\{ \left[ \left(\tilde{A}_\mu^x\right)^2 + \left(\tilde{A}_\mu^y\right)^2 \right]
b 
\right. \nonumber \\ && \ \ \ \ \ \ \ \ \ \ \ \ \ \ \ \ \ \ \ \ \
- \left. \left(\tilde{A}_\mu^z\right)^2 2 b (1+b) \right\}  C^x({\bf 0}),
\nonumber \\
-\left[ \left<H_1 H_{c2}^2 H_{c4}\right>_{\varphi c} H_{c0} \right]_D
&=& 
-2 \lambda^2 t_\mu^{-2} t_s^{-1}  \int d^2{\bf x} 
\left\{ \left[ \left(\tilde{A}_\mu^x\right)^2 + \left(\tilde{A}_\mu^y\right)^2 \right]
(1+b)(2+b)\right. \nonumber \\ && \ \ \ \ \ \ \ \ \ \ \ \ \ \ \ \ \ \ \ \ \
+ \left. \left(\tilde{A}_\mu^z\right)^2 2 (1+b) (1-b)^2 \right\}  C^x({\bf 0}),
\nonumber \\
\left[ \left<H_{c1} H_{c2}\right>_{\varphi c}^2\right]_D
&=& 
2 \lambda^2 t_\mu^{-2} t_s^{-1}  \int d^2{\bf x} 
\left\{ \left[ \left(\tilde{A}_\mu^x\right)^2 + \left(\tilde{A}_\mu^y\right)^2 \right]
b^2 \right. \nonumber \\ && \ \ \ \ \ \ \ \ \ \ \ \ \ \ \ \ \ \ \ \ \
+ \left. \left(\tilde{A}_\mu^z\right)^2 (2 + t_s t_\mu^{-1})b^2  \right\}
  C^x({\bf 0}), \nonumber \\
\frac{1}{4}\left[ \left<H_{1} H_{c2}^2\right>_{\varphi c}^2\right]_D
&=& \lambda^2 t_\mu^{-1} t_s^{-2}  \int d^2{\bf x} 
\left\{ \left[ \left(\tilde{A}_\mu^x\right)^2 + \left(\tilde{A}_\mu^y\right)^2 \right]
(1+b)^2  \right. \nonumber \\ && \ \ \ \ \ \ \ \ \ \ \ \ \ \ \ \ \ \ \ \ \
+ \left. \left(\tilde{A}_\mu^z\right)^2 (1-b)^2  \right\}
  C^x({\bf 0}), \nonumber \\
-\left[ \left<H_{1} H_{c2}^2\right>_{\varphi c}
\left<H_{c1} H_{c2}\right>_{\varphi c}\right]_D
&=& -2 \lambda^2 t_\mu^{-2} t_s^{-1}  \int d^2{\bf x} 
\left\{ \left[ \left(\tilde{A}_\mu^x\right)^2 + \left(\tilde{A}_\mu^y\right)^2 \right]
b (1+b)  \right. \nonumber \\ && \ \ \ \ \ \ \ \ \ \ \ \ \ \ \ \ \ \ \ \ \
+ \left.
\left(\tilde{A}_\mu^z\right)^2 2 b (b-1) \right\}  C^x({\bf 0}).
\nonumber 
\end{eqnarray}
The sum of the above terms is (we now again set $t_\mu=t_s$) 
\begin{eqnarray}
\label{rendis}
\lambda t_s^{-2} \int d^2{\bf x} \left\{
\left[ \left(\tilde{A}_\mu^x\right)^2 + \left(\tilde{A}_\mu^y\right)^2 \right] 
\left( 1 + \frac{ 4 (b-1) t_s + (b^2-3) \lambda}{t_s} C^x({\bf 0})
\right) \right. \nonumber \ \ \ \ \ \ \ \\  
\left. 
+ \left(\tilde{A}_\mu^z\right)^2 
\left(  (1+b)^2 - \frac{4 (1+b)^3 t_s + (3+6b+b^2)\lambda}{t_s}
 C^x({\bf 0}) \right) 
\right\}.
\end{eqnarray}

\subsubsection{On the calculation of  disorder terms}

As an illustration, 
we give details for the calculation of the variance terms for
a relatively simple term, $\left[ \left< H_{c3} \right>_{\varphi c} H_{c0}
\right]_D$, and a more involved one, $\left[ \left< H_{c1} H_{c2}^2 H_{c4}
\right>_{\varphi c} H_{c0}\right]_D$. For
$\left[ \left< H_{c3} \right>_{\varphi c} H_{c0} \right]_D$ we have
\begin{eqnarray}
\left[ \left< H_{c3} \right>_{\varphi c} H_{c0} \right]_D &=&
8 \int d^2{\bf x} d^2{\bf x}^\prime p_{k \mu} p_{k^\prime \mu^\prime}
\epsilon_{ijk} \epsilon_{i^\prime j^\prime k^\prime}
\epsilon_{abc} \epsilon_{a^\prime b^\prime c^\prime} C^l({\bf 0})
 \tilde{n}_{j^\prime}^{a^\prime} \tilde{n}_{k^\prime}^{c^\prime} 
\tilde{A}_{\mu^\prime}^{i^\prime} \times \nonumber \\
&& \left\{ 2 \epsilon_{jlm} \epsilon_{klq} \tilde{n}_m^a \tilde{n}_q^c 
\tilde{A}_\mu^i 
+ 2 \epsilon_{ilm} \epsilon_{klq} \tilde{n}_q^c \tilde{n}_j^a 
\tilde{A}_\mu^m +
2 \epsilon_{ilm}\epsilon_{jlq}\tilde{n}_q^a \tilde{n}_k^c \tilde{A}_\mu^m 
\right. \nonumber \\ && \left.
- \tilde{n}_j^a \tilde{n}_k^c \tilde{A}_\mu^i
 \left( (\epsilon_{ilm})^2 + (\epsilon_{jlm})^2
+ (\epsilon_{klm})^2 \right)  
 \right\} \left[ Q_\mu^b({\bf x}) Q_{\mu^\prime}^{b^\prime}({\bf x}^\prime)
\right]_D  
\nonumber \\
&=& 16 \lambda \int d^2{\bf x}  p_{k \mu} p_{k^\prime \mu}
\epsilon_{ijk} \epsilon_{i^\prime j^\prime k^\prime}
\epsilon_{abc} \epsilon_{a^\prime b c^\prime} 
\tilde{n}_{j^\prime}^{a^\prime} \tilde{n}_{k^\prime}^{c^\prime} 
\tilde{A}_{\mu}^{i^\prime} \times \nonumber \\ &&
\left\{
 \left( 
 \epsilon_{jlm} \epsilon_{klq} \tilde{n}_m^a \tilde{n}_q^c 
\tilde{A}_\mu^i 
+  \epsilon_{ilm} \epsilon_{klq} \tilde{n}_q^c \tilde{n}_j^a 
\tilde{A}_\mu^m +
 \epsilon_{ilm}\epsilon_{jlq}\tilde{n}_q^a \tilde{n}_k^c \tilde{A}_\mu^m 
\right) C^l({\bf 0})
\nonumber \right. \\ 
&& 
\left. 
-\tilde{n}_j^a \tilde{n}_k^c \tilde{A}_\mu^i \left( 2 + (1+b)^{-1}
\right) C^x(\bf 0)
\right\} \nonumber \\
&=& 16 \lambda \int d^2{\bf x}  \left( \epsilon_{ijk} \right)^2
 \left( \tilde{A}_\mu^i \right)^2
\left\{ p_{k\mu} p_{j\mu} C^i({\bf x}) +  p_{k\mu}^2 C^i({\bf x}) 
+ p_{k\mu} p_{i\mu} C^j({\bf x})  \nonumber \right. \\
&& \left. +
p_{k\mu} p_{i\mu} C^k({\bf x})
+ p_{k\mu}^2 C^k({\bf x}) +p_{k\mu} p_{j\mu} C^k({\bf x}) 
\right. \nonumber \\ && \left.
- \left( 2 + (1+b)^{-1} \right) C^x({\bf 0})
\left(p_{k\mu}^2 + p_{k\mu}p_{j\mu} \right)
\right\},
\end{eqnarray}
where we again used the orthonormality of the ${\bf n}_k$. Performing
the summation over the silent indices, one finally obtains
\begin{eqnarray}
&=&
16 \lambda \int d^2 {\bf x} \left\{ \left[\left(\tilde{A}_\mu^x\right)^2
+ \left(\tilde{A}_\mu^y\right)^2 \right] \left( p_{1\mu}^2 -p_{3\mu}^2
\right) + \left(\tilde{A}_\mu^z\right)^2 \left(4 p_{3\mu} p_{1\mu}-4p_{1\mu}^2
\right) \right\} C^x({\bf 0}) \nonumber \\
&=&
4 \lambda t_\mu^{-2} 
\int d^2{\bf x} \left\{ \left[ \left(\tilde{A}_\mu^x\right)^2
+ \left(\tilde{A}_\mu^y\right)^2 \right] b - 
\left(\tilde{A}_\mu^z\right)^2 
2 b (1+b) \right\} C^x({\bf 0}). \nonumber
\end{eqnarray}
We now turn to the more lengthy evaluation of  
$\left[ \left< H_{1} H_{c2}^2 H_{c4}
\right>_{\varphi c} H_{c0}\right]_D$. We have
\begin{eqnarray}
\label{longone}
H_{1}H_{c2}^2 H_{c4}&=& 16 \int d^2{\bf x} d^2{\bf x}^\prime 
d^2{\bf x}^{\prime\prime}d^2{\bf x}^{\prime\prime\prime}
p_{k\mu} t_{\mu^\prime}^{-1} p_{k^\prime \mu^\prime}
p_{k^{\prime\prime} \mu^{\prime\prime}}
p_{k^{\prime\prime\prime} \mu^{\prime\prime\prime}}
\epsilon_{ijk} \epsilon_{i^\prime j^\prime k^\prime}
\epsilon_{i^{\prime\prime} j^{\prime\prime} k^{\prime\prime}}
\epsilon_{i^{\prime\prime\prime} j^{\prime\prime\prime} 
k^{\prime\prime\prime}} \nonumber \times \\
&&\epsilon_{abc}\epsilon_{a^\prime b^\prime c^\prime}
\epsilon_{a^{\prime\prime} b^{\prime\prime} c^{\prime\prime}}
\epsilon_{a^{\prime\prime\prime} b^{\prime\prime\prime} c^{\prime\prime\prime}}
\tilde{A}_{\mu^\prime}^{i^\prime}
\left(1-b \delta_{i^\prime z}+ 2b \delta_{j^\prime z}\right)   
\tilde{n}_{j^{\prime \prime}}^{a^{\prime\prime}}
\tilde{n}_{k^{\prime \prime}}^{c^{\prime\prime}}
\tilde{n}_{j^{\prime \prime\prime}}^{a^{\prime\prime\prime}}
\tilde{n}_{k^{\prime \prime\prime}}^{c^{\prime\prime\prime}} 
\times \nonumber \\
&&  \left\{ 2 \partial_\mu \varphi^i \varphi^d
\left( \epsilon_{kdl}\tilde{n}_l^c \tilde{n}_j^a + 
\epsilon_{jdl}\tilde{n}_l^a \tilde{n}_k^a \right)
+ \varphi^d \partial_\mu \varphi^l \epsilon_{idl}\tilde{n}_j^a\tilde{n}_k^c
\right\} \nonumber \times \\ &&
\partial_{\mu^\prime}\varphi^{j^\prime} \varphi^{k^\prime}
\partial_{\mu^{\prime\prime}}\varphi^{i^{\prime\prime}}
\partial_{\mu^{\prime\prime\prime}}\varphi^{i^{\prime\prime\prime}}
Q_{\mu}^b Q_{\mu^{\prime\prime}}^{b^{\prime\prime}}
Q_{\mu^{\prime\prime\prime}}^{b^{\prime\prime\prime}}.
\end{eqnarray}
We now need to perform the average over the $\varphi$ fields. 
For convenience, we split $H_{1}H_{c2}^2 H_{c4}={\cal A}+ {\cal B}$
into two terms,
where ${\cal A}$ corresponds to the part of $H_{1}H_{c2}^2 H_{c4}$ which
involves the first term in the
curly brackets in Eq.~(\ref{longone}) and  ${\cal B}$ corresponds
to the second term in the curly brackets.
For $\left<{\cal A}\right>_{\varphi c}$, we need to calculate the average
\begin{eqnarray}
\left< \partial_\mu \varphi^i ({\bf x}) \varphi^d({\bf x})
 \partial_{\mu^\prime} \varphi^{j^\prime} ({\bf x}^\prime) 
\varphi^{k^\prime}({\bf x}^\prime)
\partial_{\mu^{\prime\prime}}\varphi^{i^{\prime\prime}}
({\bf x}^{\prime \prime})
\partial_{\mu^{\prime\prime\prime}}\varphi^{i^{\prime\prime\prime}}
({\bf x}^{\prime\prime\prime})
\right>_{\varphi},
\end{eqnarray}
which can be easily done via  Wick's Theorem. However, not all
possible permutations of pairings will contribute. All terms
involving either of the contractions $\left<id\right>$ or
$\left<j^\prime k^\prime \right>$ vanish as $\partial_\mu C^x({\bf 0})=0$.
Although not immediately apparent, terms involving the pairing
$\left< i^{\prime \prime} i^{\prime\prime\prime} \right>$ also
do not contribute to one loop order. This can be seen only
after the computation of the disorder average $\left[ \left<
{\cal A} \right>_{\varphi c} H_{c0}\right]_D$ and a gradient expansion similar
to the one employed below Eq.~(\ref{zeroterm}). Using the same arguments
as we used for the term (\ref{zeroterm}), all 
$\left< i^{\prime \prime} i^{\prime\prime\prime} \right>$ contractions
can then be shown to give no contribution. Furthermore, all contractions
which are identical up to a permutation of the indices $i^{\prime\prime}$
and $i^{\prime \prime \prime}$ will give the same contributions 
after the disorder average is taken, as discussed below. We 
therefore only write down half of the permutations and indicate
the others by 
$\left\{ \prime\prime\leftrightarrow \prime \prime \prime\right\}$.
Thus, we only need to
keep the following terms,
\begin{eqnarray}
\label{phiterms}
&& \left< \partial_\mu \varphi^i ({\bf x}) \varphi^d({\bf x})
 \partial_{\mu^\prime} \varphi^{j^\prime} ({\bf x}^\prime) 
\varphi^{k^\prime}({\bf x}^\prime)
\partial_{\mu^{\prime\prime}}\varphi^{i^{\prime\prime}}
({\bf x}^{\prime \prime})
\partial_{\mu^{\prime\prime\prime}}\varphi^{i^{\prime\prime\prime}}
({\bf x}^{\prime\prime\prime})
\right>_{\varphi} \rightarrow \nonumber \\ 
&&\ \ \ \ \ \ \ \ \ \ \ \ \ \
  \delta_{ij^\prime}\delta_{d i^{\prime\prime}}
\delta_{k^\prime i^{\prime\prime\prime}}
\partial_{\mu}\partial_{\mu^\prime}C^i({\bf x}-{\bf x}^\prime)
\partial_{\mu^{\prime \prime}} C^d({\bf x}-{\bf x}^{\prime\prime})
\partial_{\mu^{\prime \prime \prime}} 
C^{i^{\prime\prime\prime}}({\bf x}^\prime-{\bf x}^{\prime\prime\prime})
\nonumber \\ & &
\ \ \ \ \ \ \ \ \ \ \ \,
+  \delta_{ik^\prime}\delta_{d i^{\prime\prime}}
\delta_{j^\prime i^{\prime\prime\prime}}
\partial_{\mu}C^i({\bf x}-{\bf x}^\prime)
\partial_{\mu^{\prime \prime}} C^d({\bf x}-{\bf x}^{\prime\prime})
\partial_{\mu^{\prime}} \partial_{\mu^{\prime \prime \prime}} 
C^{i^{\prime\prime\prime}}({\bf x}^\prime-{\bf x}^{\prime\prime\prime})
\nonumber \\ & &
\ \ \ \ \ \ \ \ \ \ \ \,
+  \delta_{ii^{\prime\prime}}\delta_{d j^{\prime}}
\delta_{k^\prime i^{\prime\prime\prime}}
\partial_{\mu}\partial_{\mu^{\prime \prime}}C^i({\bf x}-{\bf x}^{\prime\prime})
\partial_{\mu^{\prime}} C^d({\bf x}-{\bf x}^{\prime})
\partial_{\mu^{\prime\prime \prime}}  
C^{i^{\prime\prime\prime}}({\bf x}^\prime-{\bf x}^{\prime\prime\prime})
\nonumber \\ & &
\ \ \ \ \ \ \ \ \ \ \ \,
+  \delta_{ii^{\prime\prime}}\delta_{d k^{\prime}}
\delta_{j^\prime i^{\prime\prime\prime}}
\partial_{\mu}\partial_{\mu^{\prime \prime}}C^i({\bf x}-{\bf x}^{\prime\prime})
C^d({\bf x}-{\bf x}^{\prime})
\partial_{\mu^{\prime}}\partial_{\mu^{\prime\prime \prime}}  
C^{i^{\prime\prime\prime}}({\bf x}^\prime-{\bf x}^{\prime\prime\prime})
\nonumber \\ & &
\ \ \ \ \ \ \ \ \ \ \ \, + \left\{ \prime\prime \leftrightarrow 
\prime\prime\prime \right\}
\end{eqnarray}
Let us now perform the disorder average 
$\left[\left<{\cal A}\right>_{\varphi c}
H_{c0} \right]_D$. For this, we need to calculate
\begin{eqnarray}
\left[ Q_{\tilde{\mu}}^{\tilde{b}}(\tilde{\bf x}) 
Q_\mu^b({\bf x}) Q_{\mu^{\prime\prime}}^{b^{\prime\prime}}
({\bf x}^{\prime\prime})
Q_{\mu^{\prime\prime\prime}}^{b^{\prime\prime\prime}}
({\bf x}^{\prime\prime\prime})  \right]_D
\end{eqnarray}
where the variables carrying a tilde arise from the $H_{c0}$ term.
Again, we can use Wick's Theorem to decompose the average. 
Of the three possible permutations of pairings, two involve either
of the two contractions $\left<b b^{\prime \prime}\right>$ or
$\left<b b^{\prime \prime \prime}\right>$. Neither permutation
contributes. This is easily seen for the $\left<b b^{\prime \prime}\right>$
contraction and the explicitly written terms in (\ref{phiterms})
because  they all involve after the contraction
a derivative of $C^x({\bf 0})$ and thus vanish. The same terms
also do not contribute for the case of a 
$\left<b b^{\prime \prime \prime}\right>$ contraction, which again 
can be seen with a gradient expansion and using arguments analogous to those
below Eq.~(\ref{zeroterm}). Therefore, only one term of the disorder average
must be kept,
\begin{eqnarray}
\label{qterms}
\left[ Q_{\tilde{\mu}}^{\tilde{b}}(\tilde{\bf x}) 
Q_\mu^b({\bf x}) Q_{\mu^{\prime\prime}}^{b^{\prime\prime}}
({\bf x}^{\prime\prime})
Q_{\mu^{\prime\prime\prime}}^{b^{\prime\prime\prime}}
({\bf x}^{\prime\prime\prime})  \right]_D \rightarrow 
\lambda^2\delta_{b\tilde{b}} \delta_{b^{\prime\prime}b^{\prime\prime\prime}}
\delta_{\mu\tilde{\mu}}\delta_{\mu^{\prime\prime}\mu^{\prime\prime\prime}}
\delta({\bf x}-\tilde{\bf x})
\delta({\bf x}^{\prime\prime}-{\bf x}^{\prime\prime\prime})
\end{eqnarray}
The terms in (\ref{phiterms})
which only differ by a permutation of the double primed and triple primed
variables give then identical contributions, as such a permutation simply 
relabels the variables associated with the two $H_{c2}$ terms in 
$\left[ \left< {\cal A} \right>_{\varphi c} H_{c0}\right]_D$.
With (\ref{phiterms}, \ref{qterms}) we then have 
\begin{eqnarray}
\left[ \left< {\cal A} \right>_{\varphi c} H_{c0}\right]_D &=& 
128 \lambda^2 \int d^2{\bf x} d^2{\bf x}^\prime d^2{\bf x}^{\prime\prime}
t_{\mu^\prime}^{-1}  \left( p_{k^{\prime \prime} \mu^{\prime \prime}}^2
+p_{k^{\prime \prime} \mu^{\prime \prime}} 
p_{j^{\prime \prime} \mu^{\prime \prime}}\right)^2 
\left(\epsilon_{i^{\prime\prime}j^{\prime\prime}k^{\prime\prime}}\right)^2
\epsilon_{i^{\prime}j^{\prime}k^{\prime}} \tilde{A}_{\mu^\prime}^{i^\prime}
\nonumber \times \\ &&
\left(1-b \delta_{i^\prime z} + 2 b \delta_{j^\prime z}\right)
\left\{-\tilde{A}_\mu^k p_{i\mu}p_{k\mu} \epsilon_{kdi} + 
\tilde{A}_\mu^k p_{d\mu}p_{k\mu} \epsilon_{idk} -
\tilde{A}_\mu^j p_{d\mu}^2 \epsilon_{ijd}  \right.
\nonumber \\ && \left.
-\tilde{A}_\mu^j p_{i\mu}p_{d\mu} \epsilon_{jdi}\right\} \times \nonumber
\\ &&
\left[  
\delta_{ij^\prime}\delta_{d i^{\prime\prime}}
\delta_{k^\prime i^{\prime\prime}}
\partial_{\mu}\partial_{\mu^\prime}C^i({\bf x}-{\bf x}^\prime)
\partial_{\mu^{\prime \prime}} C^d({\bf x}-{\bf x}^{\prime\prime})
\partial_{\mu^{\prime \prime}} 
C^{i^{\prime\prime}}({\bf x}^\prime-{\bf x}^{\prime\prime})
\right.
\nonumber \\ && 
+  \delta_{ik^\prime}\delta_{d i^{\prime\prime}}
\delta_{j^\prime i^{\prime\prime}}
\partial_{\mu}C^i({\bf x}-{\bf x}^\prime)
\partial_{\mu^{\prime \prime}} C^d({\bf x}-{\bf x}^{\prime\prime})
\partial_{\mu^{\prime}} \partial_{\mu^{\prime \prime}} 
C^{i^{\prime\prime}}({\bf x}^\prime-{\bf x}^{\prime\prime})
\nonumber \\ & &
+  \delta_{ii^{\prime\prime}}\delta_{d j^{\prime}}
\delta_{k^\prime i^{\prime\prime}}
\partial_{\mu}\partial_{\mu^{\prime \prime}}C^i({\bf x}-{\bf x}^{\prime\prime})
\partial_{\mu^{\prime}} C^d({\bf x}-{\bf x}^{\prime})
\partial_{\mu^{\prime \prime}}  
C^{i^{\prime\prime}}({\bf x}^\prime-{\bf x}^{\prime\prime})
\nonumber \\ & & \left.
+  \delta_{ii^{\prime\prime}}\delta_{d k^{\prime}}
\delta_{j^\prime i^{\prime\prime}}
\partial_{\mu}\partial_{\mu^{\prime \prime}}C^i({\bf x}-{\bf x}^{\prime\prime})
C^d({\bf x}-{\bf x}^{\prime})
\partial_{\mu^{\prime}}\partial_{\mu^{\prime \prime}}  
C^{i^{\prime\prime}}({\bf x}^\prime-{\bf x}^{\prime\prime})
\right].
\end{eqnarray}
The integration over ${\bf x}^{\prime \prime}$ can now be
performed with 
\begin{eqnarray}
\label{intx}
t_{\mu^{\prime \prime}}^{-1} \int d^2{\bf x}^{\prime \prime}
\partial_{\mu^{\prime \prime}} C^x({\bf x}-{\bf x}^{\prime\prime})
\partial_{\mu^{\prime \prime}} C^x({\bf x}^\prime -{\bf x}^{\prime\prime})
=\frac{1}{2} C^x({\bf x}-{\bf x}^\prime) + {\cal O}(\kappa^2).
\end{eqnarray} 
The remaining double integral over ${\bf x}$ and ${\bf x}^\prime$ can
then again be approximated with a gradient expansion in the relative
coordinate and employing Eq.~(\ref{intx}). We then obtain (we denote the 
center of mass coordinate again by ${\bf x}$)
\begin{eqnarray}
\left[ \left< {\cal A} \right>_{\varphi c} H_{c0}\right]_D &\simeq& 
16 \lambda^2 t_s \int d^2{\bf x} 
\left( p_{k^{\prime \prime}}^2
+p_{k^{\prime \prime}} 
p_{j^{\prime \prime}}\right)^2 
\left(\epsilon_{i^{\prime\prime}j^{\prime\prime}k^{\prime\prime}}\right)^2
\epsilon_{i^{\prime}j^{\prime}k^{\prime}} \tilde{A}_{\mu}^{i^\prime}
\left(1-b \delta_{i^\prime z} + 2 b \delta_{j^\prime z}\right)
\times
\nonumber \\ &&
\beta_i \beta_d
\epsilon_{idk}
\left\{\tilde{A}_\mu^k p_{i\mu}p_{k\mu} + 
\tilde{A}_\mu^k p_{d\mu}p_{k\mu} +
\tilde{A}_\mu^j p_{d\mu}^2 
+\tilde{A}_\mu^j p_{i\mu}p_{d\mu}\right\} \times \nonumber
\\ &&
\left[  
\delta_{ij^\prime}\delta_{d i^{\prime\prime}}
\delta_{k^\prime i^{\prime\prime}}
\beta_{k^\prime}
-  \delta_{ik^\prime}\delta_{d i^{\prime\prime}}
\delta_{j^\prime i^{\prime\prime}}
\beta_{j^\prime}
- \delta_{ii^{\prime\prime}}\delta_{d j^{\prime}}
\delta_{k^\prime i^{\prime\prime}}
\beta_{k^\prime}
+  \delta_{ii^{\prime\prime}}\delta_{d k^{\prime}}
\delta_{j^\prime i^{\prime\prime}}
\beta_{j^\prime}
\right] \times  \nonumber
\\ && \times C^x({\bf 0}).   
\end{eqnarray}
where $\beta_k$ is defined through $\beta_1=\beta_2=1$, 
$\beta_3=(1+b)^{-1}$ and $p_{k} t_s /t_\mu=p_{k \mu}$. After some
straightforward algebra, one finally finds
\begin{eqnarray} 
\left[ \left< {\cal A} \right>_{\varphi c} H_{c0}\right]_D &\simeq& 
32 \lambda^2 t_\mu^{-2} t_s^3 \int d^2{\bf x} \left\{
\left[ \left( \tilde{A}_\mu^x\right)^2 +\left( \tilde{A}_\mu^y\right)^2\right]
\times \right. \nonumber \\ && \left. \ \ \ \
\left( \left( p_1+p_3\right)^2 + \frac{4 p_1^2}{1+b} \right)
\left( \left(p_1 + p_3\right)^2 + 2 p_1^2 + 2 p_1 p_3 \right) \right.
\nonumber + \\ && \left. \ \ \ \
\left( \tilde{A}_\mu^z\right)^2 8 (1-b) \left(p_1^2 + p_1 p_3 \right)
\left(p_1 + p_3\right)^2 \right\} C^x({\bf 0}).  
\end{eqnarray}
The calculation of $\left[ \left< {\cal B}\right>_{\varphi c} H_{c0}\right]_D$
can be done in much the same way as just shown for
$\left[ \left< {\cal A}\right>_{\varphi c} H_{c0}\right]_D$. One arrives at
\begin{eqnarray}
\left[ \left< {\cal B} \right>_{\varphi c} H_{c0}\right]_D &\simeq& 
-32 \lambda^2 t_\mu^{-2} t_s^3 \int d^2{\bf x} \left\{
\left[ \left( \tilde{A}_\mu^x\right)^2 +\left( \tilde{A}_\mu^y\right)^2\right]
\left( \left(p_1 + p_3\right)^2 + \frac{4 p_1^2}{1+b} \right) 
\left( p_1+p_3\right)^2 \right.
\nonumber \\ && \ \ \ \ \ \ \ \ \ \ \ \ \ \ \ \ \ \ + \left. 
\left( \tilde{A}_\mu^z\right)^2 8 (1-b) p_1^2 
\left(p_1 + p_3\right)^2 \right\} C^x({\bf 0}).  
\end{eqnarray}
Finally, expressing all $p_k$ through $b$ and $t_s$, one obtains for
$\left[ \left< {\cal A}+{\cal B} \right>_{\varphi c}H_{c0} \right]_D$
\begin{eqnarray}
\left[ \left< H_1 H_{c2}^2 H_{c4} \right>_{\varphi c} H_{c0} \right]_D
&=&
2 \lambda^2 t_\mu^{-2} t_s^{-1}  \int d^2{\bf x} 
\left\{ \left[ \left(\tilde{A}_\mu^x\right)^2 + 
\left(\tilde{A}_\mu^y\right)^2 \right]
(1+b)(2+b)\right. \nonumber \\ && \ \ \ \ 
+ \left. \left(\tilde{A}_\mu^z\right)^2 2 (1+b) (1-b)^2 \right\}  C^x({\bf 0}).
\end{eqnarray}

\end{widetext}
\end{appendix}

\end{document}